\documentclass[twocolumn,a4paper,showpacs,pra,aps]{revtex4-1}

 \usepackage{amsmath}
 \usepackage{amssymb}
 \usepackage{bbold}
 \usepackage{latexsym}
 \usepackage{amsfonts}
 \usepackage[caption=false]{subfig}
 \usepackage{epsfig}
 \usepackage{psfrag}
 \usepackage{color}
 \usepackage{overpic}
 \definecolor{darkblue}{rgb}{0,0,.5}
 \usepackage[linktocpage, colorlinks=true ,linkcolor=darkblue, citecolor=darkblue]{hyperref}
 \usepackage[all]{hypcap}

 \newcommand{\ket}[1]{\left|#1\right>}
 \newcommand{\bra}[1]{\left<#1\right|}
 \newcommand{\expval}[1]{\left< #1 \right>}
 
 \newcommand{\braket}[2]
 {\left<#1|#2\right>}
 \newcommand{\nn}{\nonumber\\}
 
 \newcommand{\f}[1]{\mbox{\boldmath$#1$}}

 \newcommand{\bea}{\begin{eqnarray}}
 \newcommand{\eea}{\end{eqnarray}}
 
 \newcommand{\trace}[1]{{\rm Tr}\left\{ #1 \right\}}
 \newcommand{\traceB}[1]{{\rm Tr_B}\left\{ #1 \right\}}
 
 \newcommand{\abs}[1]{{\left| #1 \right|}}
 
 \newcommand{\HS}{\mathcal{H}_{\rm S}}
 \newcommand{\HB}{\mathcal{H}_{\rm B}}
 \newcommand{\HI}{\mathcal{H}_{\rm SB}}

 \newcommand{\ii}{{\rm i}}
\begin{document}

\title{Transport as a sensitive indicator of quantum criticality}

\author{Gernot Schaller}\email{gernot.schaller@tu-berlin.de}
\author{Malte Vogl}
\author{Tobias Brandes}

\affiliation{Institut f\"ur Theoretische Physik, Technische Universit\"at Berlin, Hardenbergstr. 36, 10623 Berlin, Germany}

\begin{abstract}
We consider bosonic transport through one-dimensional spin systems.
Transport is induced by coupling the spin systems to bosonic reservoirs kept at different
temperatures.
In the limit of weak-coupling between spins and bosons we apply the quantum-optical master equation
to calculate the energy transmitted from source to drain reservoirs.
At large thermal bias, we find that the current for longitudinal transport becomes 
independent of the chain length and is also not drastically affected by the presence of disorder.
In contrast, at small temperatures, the current scales inversely with the chain length and 
is further suppressed in presence of disorder.
We also find that the critical behaviour of the ground state is mapped to critical behaviour of the current
-- even in configurations with infinite thermal bias.
\end{abstract}


\maketitle


\section{Introduction}

Quantum phase transitions are drastic changes of a system's ground state when an
external control parameter is smoothly varied across a critical value~\cite{sachdev2000}.
Typically, they occur in the continuum limit where the considered system becomes infinitely large.
Naturally, a critical behavior of the ground state leads to non-analytic behaviour of almost all observables at zero temperature.
However, realistic systems cannot be kept at zero temperature nor can they be perfectly isolated from
their environments.
In such more generalized scenarios involving dissipation, feedback control or external driving, 
excited states -- which may also exhibit critical behaviour~\cite{cejnar2006a,caprio2008a,perez_fernandez2011a,dietz2013a,brandes2013a} -- become relevant.
 
On the one hand, such modifications may arise as parts of a detecting environment, as is e.g. exemplified by recent advances
in cold atom experiments~\cite{lignier2007a,baumann2010a,baumann2011a}.
On the other hand, there is evidence that additional couplings may give rise to even richer phase diagrams both in case
of driving~\cite{inoue2010a,lindner2011a,jiang2011a,bastidas2012a,bastidas2012b} or dissipation~\cite{morrison2008a,dalla_torre2010a,bhaseen2012a,kessler2012a,hoening2012a}.

In contrast, here we will explore the scenario of heat transport through spin chains in the weak-coupling limit, for several reasons:
First, in this limit, we do not expect the phase diagram to be modified, and the heat current may rather carry signatures of
the critical model behaviour~\cite{lambert2009a,vogl2012b}.
Second, we note that the heat transferred between a quantum system and a reservoir may be unambiguously defined 
in the weak-coupling limit~\cite{esposito2010a}, whereas this becomes an intricate issue on its own beyond~\cite{campisi2011a}.
Third, this limit can be conveniently described by master equations of Lindblad type that preserve positivity~\cite{lindblad1976a} and
obey detailed balance relations that induce thermalization with the (fixed) reservoir temperature in equilibrium setups~\cite{lindblad1976a}.
Fourth, we consider Ising-type spin chains as these can be easily diagonalized.
In particular the quantum Ising model in a transverse field has been an attractive candidate of
theoretical studies, and several proposals for its experimental implementation with various systems 
exist~\cite{friedenauer2008a,mostame2008a,zhang2009a,coldea2010a,edwards2010a,kim2011b}.
Experimental conditions can usually hardly be perfectly controlled, such that we will also consider the impact of
disorder on the heat transport.

A non-equilibrium setup may be engineered by connecting the system to reservoirs at different temperature, 
which induces for two terminals a heat current from source to drain.
We note that basic laws of thermodynamics should be respected even in far from equilibrium setups.
For example, the heat current should always flow from hot to cold reservoir and 
the entropy production in the system should be positive.
We note that the method we use -- 
the master equation in positivity-preserving secular approximation 
applied to multiple reservoirs -- has all these features.

This paper is organized as follows:
In Sec.~\ref{SEC:model} we introduce the physical models we consider here in detail.
In Sec.~\ref{SEC:methods} we describe our methods, namely the quantum master equation and the extraction of 
heat transport characteristics from it.
We also discuss how to diagonalize the central spin systems with or without disorder and 
review the reduced dynamics at low temperatures.
We provide heat transport characteristics for perpendicular and longitudinal 
transport through a closed spin chain in Secs.~\ref{SEC:results1} and~\ref{SEC:results2}, respectively.
Results for longitudinal transport through an open spin chain are presented in Sec.~\ref{SEC:results3}.
We close with a note on Fourier's law in Sec.~\ref{SEC:fourier} and conclusions.


\section{Model}\label{SEC:model}

We want to study heat transport between two bosonic reservoirs ($\hbar=1$ throughout)
\bea
\HB^{S} = \sum_k \omega_k^S b_{kS}^\dagger b_{kS}\,,\qquad
\HB^{D} = \sum_k \omega_k^D b_{kD}^\dagger b_{kD}
\eea
with $b_{k\alpha}$ being a bosonic annihilation operator of a particle
with frequency $\omega_k^\alpha$ in either source ($\alpha=S$) or drain ($\alpha=D$).
A thermal gradient between the reservoirs is induced by keeping them at separate
thermal equilibrium states $\rho_\alpha = e^{-\beta_\alpha \HB^\alpha}/Z_\alpha$, where
$\beta_\alpha$ denotes the inverse temperature of reservoir $\alpha$.
Without loss of generality we assume that the temperature of reservoir $S$ is larger
than the temperature of reservoir $D$ ($\beta_S < \beta_D$).

These reservoirs only interact indirectly via the exchange of energy with a spin chain composed
of $N$ spins, where we consider systems of the type
\bea
\HS &=& \sum_{i=1}^N g_i \sigma^x_i
+ \sum_{i=1}^N \left[J^y_i \sigma^y_i \sigma^y_{i+1} + J^z_i \sigma^z_i \sigma^z_{i+1}\right]\,.
\eea
Here, $\sigma^\alpha_i$ denotes the Pauli matrix $\sigma^\alpha$ acting only on the $i$-th spin
and we use the convention $\sigma^\alpha_{N+1} \equiv \sigma^\alpha_1$.
The coefficients $g_i$ denote the strength of a local external field whereas $J^{y/z}_i$ model a 
ferromagnetic ($J^{y/z}_i<0$) or anti-ferromagnetic  ($J^{y/z}_i>0$) next-neighbor interaction 
between the spins.
Specific cases are the open disordered spin chain ($J^y_N = J^z_N = 0$), 
the open YZ model ($J^y_N=J^z_N=0$, $g_i=g$, $J^y_{i<N} = J^y$, $J^z_{i<N} = J^z$), 
and the closed YZ-model ($g_i=g$, $J^y_i = J^y$, $J^z_i = J^z$), 
which further reduces for $J^y=0$ to the Ising chain in a transverse field.
These spin chains have the advantage that they can be mapped to non-interacting fermions
\bea
\HS = \sum_q \epsilon_q \eta_q^\dagger \eta_q
\eea
with fermionic annihilation operators $\eta_q$ of quasiparticles with quasienergy $\epsilon_q$.
Remarkably, this mapping can be performed with an effort that scales at most polynomially in the system size $N$~\cite{bunder1999a}
and -- sufficient symmetry provided -- can even be performed analytically~\cite{pfeuty1970a}.
Despite their simplicity, such spin chains display rich behaviour such as quantum criticality in the continuum limit $N\to\infty$, 
where the ground state (and associated observables) changes non-analytically as the parameters of the Hamiltonian are varied 
across a critical point~\cite{sachdev2000}.

\begin{figure}[ht]
\begin{center}
\includegraphics[width=0.48\textwidth]{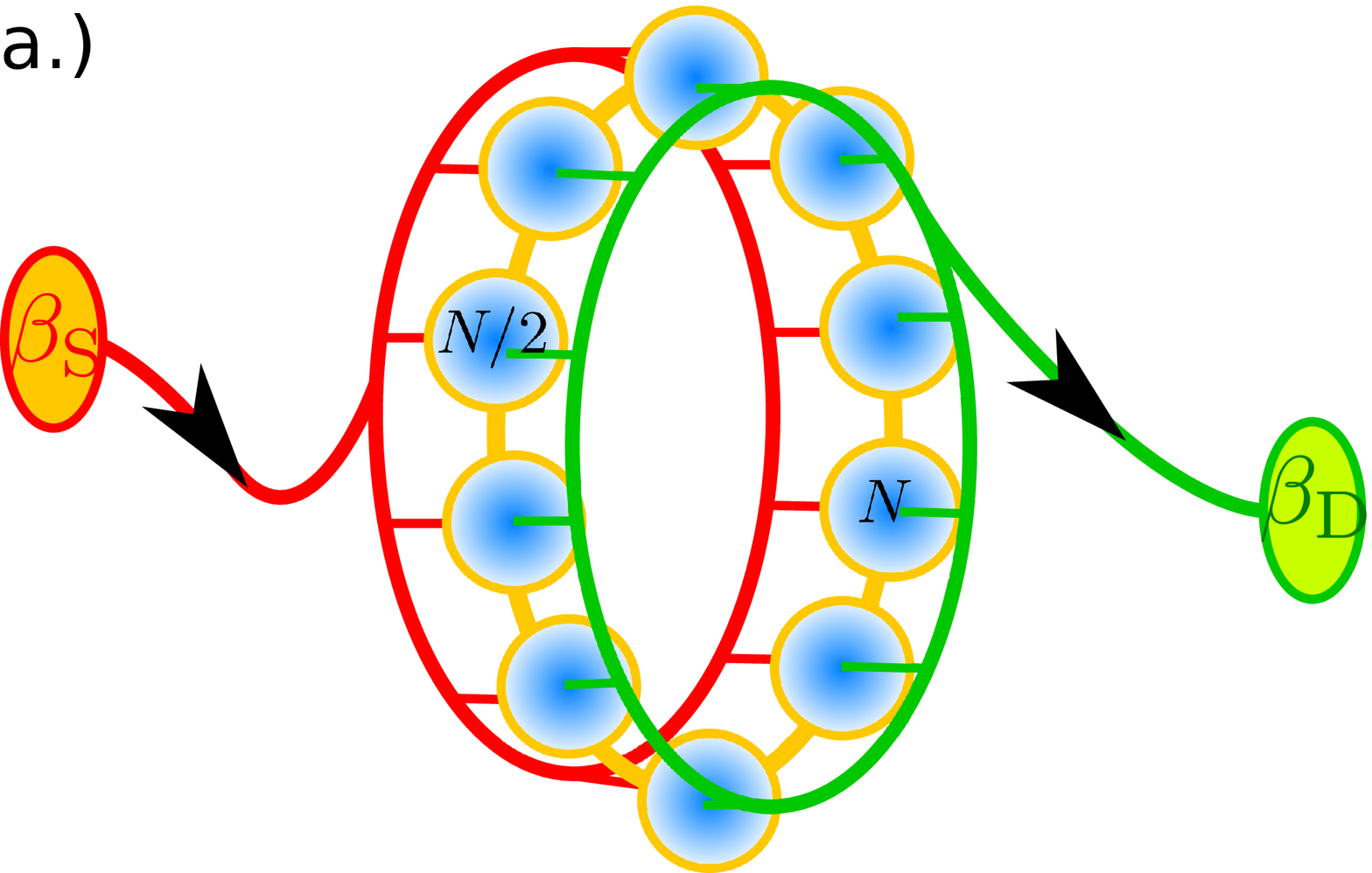}\\
\includegraphics[width=0.48\textwidth]{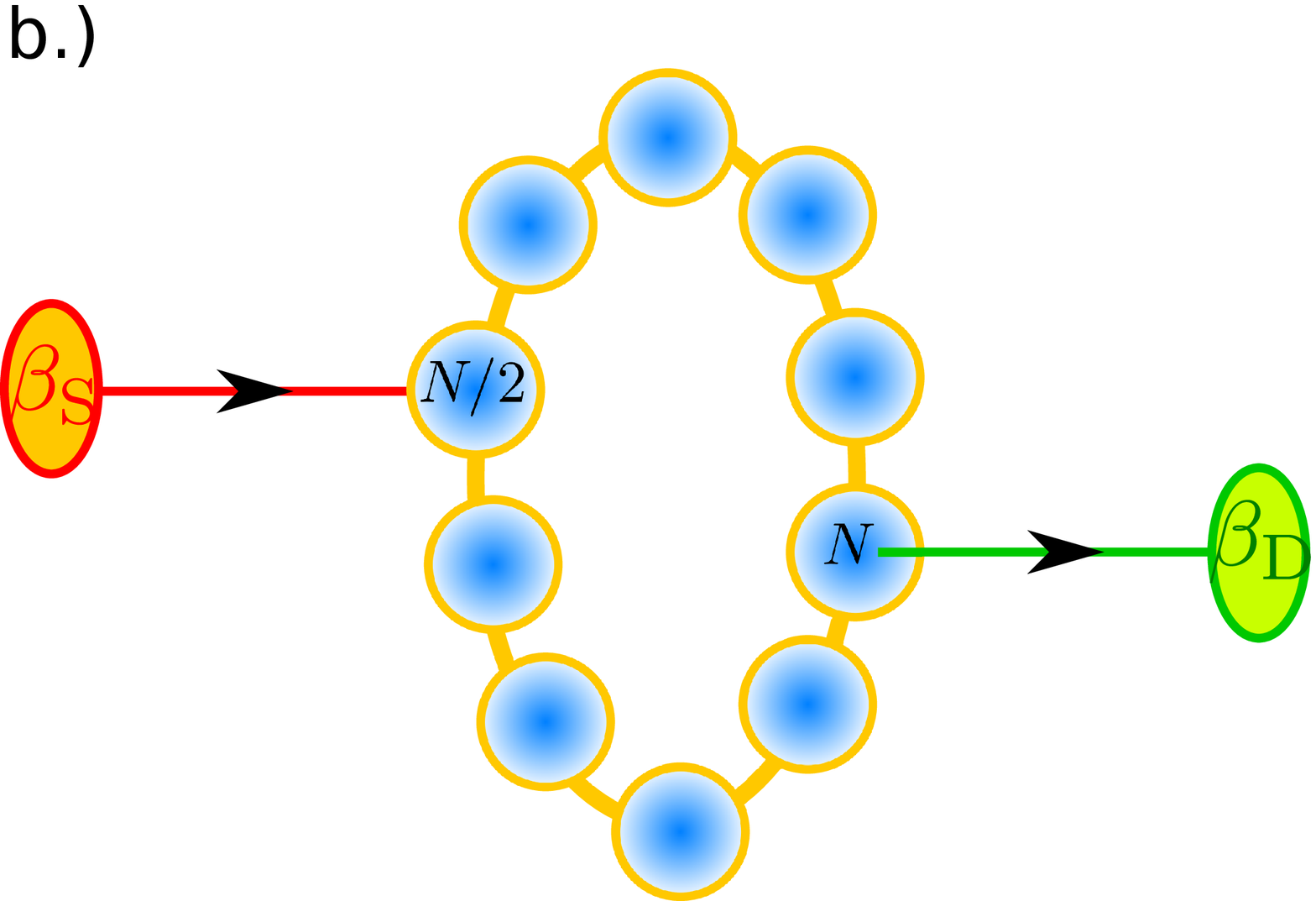}\\
\includegraphics[width=0.48\textwidth]{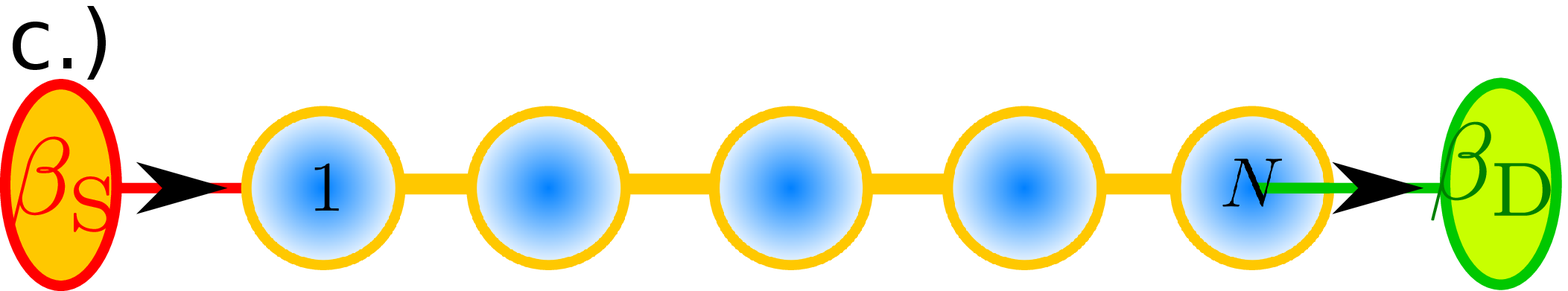}
\end{center}
\caption{\label{FIG:coupling}(Color Online)
Sketch of heat transport setups considered in this paper.
Arrows indicate the energy flow provided a temperature gradient ($\beta_S < \beta_D$) is present.
{\bf a.)} A homogeneous coupling via $J^x=\sum_i \sigma^x_i$ has been considered in~\cite{vogl2012b}.
For a homogeneous Ising chain it admits an analytic calculation of the heat current, and the results
therein may be readily generalized to the homogeneous YZ-model.
{\bf b.)}
In particular for large $N$ it is more reasonable to consider local couplings e.g. via anti-podal points, which
we can take as $N/2$ and $N$ via the coupling operator $\sigma^x_{N/2}$ and $\sigma^x_N$, respectively.
{\bf c.)}
An open spin chain can be coupled at its ends via the operators $\sigma^x_1$ and $\sigma^x_N$ to the 
bosonic reservoirs.
}
\end{figure}

There are many ways of coupling the source and drain reservoirs via spin chains, see Fig.~\ref{FIG:coupling}.
Our previous study (top panel) revealed that the heat current exhibits features of the critical ground state
behaviour even at finite temperatures and large temperature gradients -- opposed to local observables such as
mean energy or magnetization densities.
Here, we would like to learn whether this feature is more generic.
More formally, the interaction between system and reservoirs can be written as
\bea
\HI = A_S \otimes B_S + A_D \otimes B_D\,,
\eea
where $A_\alpha$ and $B_\alpha$ denote system and reservoir operators, respectively.
We consider 
\bea
B_\alpha = \sum_k \left[h_{k\alpha} b_{k\alpha} + h_{k\alpha}^* b_{k\alpha}^\dagger\right]
\eea
throughout this paper, where the $h_{k\alpha}$ ($h_{k\alpha}^*$) are the microscopic amplitudes
for the annihilation (creation) for a boson of mode $k$ in reservoir $\alpha$.
For the system contribution, we had $A_S = A_D = J^x = \sum_{i=1}^N \sigma^x_i$ in Ref.~\cite{vogl2012b} (top panel).
Here, we will consider the local couplings $A_S = \sigma^x_{N/2}$ and $A_D = \sigma^x_N$ (middle panel)
for the closed spin chain
and $A_S=\sigma^x_1$ and $A_D=\sigma^x_N$ for the open spin chain (bottom panel).

We note that although both $\HS$ and $\HB$ are diagonalizable, the presence of the interaction $\HI$ requires
a perturbative treatment.
In contrast, reservoirs consisting of spin chains too would in some cases enable an exact solution~\cite{arrachea2009a}.


\section{Methods}\label{SEC:methods}


\subsection{Master equation}

When the interaction Hamiltonian $\HI$ is much smaller than the system and reservoir parts, 
conventional perturbation theory to second order in $h_{k\alpha}$ with standard approximations eventually leads to a closed 
master equation of Lindblad form~\cite{lindblad1976a}, which in superoperator notation can 
be written as $\dot{\rho} = {\cal L} \rho$.
Typically, such a master equation is found valid at high temperatures and/or small coupling strengths.
We note that for our model $\expval{B_\alpha} = \traceB{B_\alpha \bar{\rho}_\alpha} = 0$, which implies that
to lowest order both reservoirs enter additively in the master equation ${\cal L} = {\cal L}_S + {\cal L}_D$.
In the energy eigenbasis of the system $\HS\ket{a} = E_a \ket{a}$, the master equation assumes the form
\bea
\dot{\rho} &=& -\ii \left[\HS + H_{\rm LS}, \rho\right]\nn
&&+\sum_{ab,cd} \gamma_{ab,cd}\left[L_{ab} \rho L_{cd}^\dagger - \frac{1}{2} \left\{L_{cd}^\dagger L_{ab}, \rho\right\}\right]\,,
\eea
where the Lindblad jumpers $L_{ab} \equiv \ket{a}\bra{b}$ trigger transitions between energy eigenstates and 
where $H_{\rm LS} = \sum_{ab} \tilde{\sigma}_{ab} \ket{a} \bra{b}$ denotes the Lamb-shift Hamiltonian.
We note here that microscopic derivations in the weak-coupling limit will -- a non-trivial system Hamiltonian $\HS$ provided -- generally
map local terms in the Hamiltonian to non-local Lindblad jumpers.
Theories just phenomenologically assuming local Lindblad terms should therefore always be cautiously checked for their thermodynamic
consistency~\cite{saito2003a,prosen2010a,sun2010a,li2011b}.

Since the Liouvillian is additively decomposable, this directly transfers to the coefficients
$\tilde{\sigma}_{ab} = \tilde{\sigma}_{ab}^S + \tilde{\sigma}_{ab}^D$ and
$\gamma_{ab,cd} = \gamma_{ab,cd}^S + \gamma_{ab,cd}^D$.
Denoting the reservoir-specific interaction by $\HI^\alpha = A_\alpha \otimes B_\alpha$, these become 
explicitly~\cite{breuer2002,schaller2008a,schaller2014}
\bea
\tilde{\sigma}_{ab}^\alpha &=& \frac{\delta_{E_b,E_a}}{2\ii} \sum_c \sigma_\alpha(E_a-E_c) \bra{a} A_\alpha \ket{c} \bra{c} A_\alpha \ket{b}\,,\nn
\gamma_{ab,cd}^\alpha &=& \delta_{E_b-E_a,E_d-E_c} \gamma_\alpha(E_b-E_a) \bra{a} A_\alpha \ket{b} \bra{c} A_\alpha \ket{d}^*\,.
\eea
Above, we have introduced the even {\bf (}$\gamma(\omega)${\bf )} and odd {\bf (}$\sigma(\omega)${\bf )} Fourier transform of the reservoir correlation function
\bea
\gamma_\alpha(\omega) &=& \int \expval{\f{B}_\alpha(\tau) B_\alpha} e^{+\ii\omega\tau} d\tau\,,\nn
\sigma_\alpha(\omega) &=& \int \expval{\f{B}_\alpha(\tau) B_\alpha} {\rm sgn}(\tau) e^{+\ii\omega\tau} d\tau\nn
&=& \frac{\ii}{\pi} {\cal P} \int \frac{\gamma_\alpha(\omega')}{\omega-\omega'} d\omega'\,.
\eea
We note that we generally have $\left[\HS,H_{\rm LS}\right]=0$, which implies that both operators can be 
simultaneously diagonalized.
For the considered bosonic reservoirs the correlation function simply reads~\cite{schaller2014}
\bea
\gamma_\alpha(\omega) = \Gamma_\alpha(\omega) \left[1+n_\alpha(\omega)\right]
\eea
with the Bose distribution
$n_\alpha(\omega) = \left[e^{\beta_\alpha \omega}-1\right]^{-1}$
and the spectral coupling density
\bea
\Gamma_\alpha(\omega) = 2\pi \sum_k \abs{h_{k\alpha}}^2 \delta(\omega-\omega_{k\alpha})
\eea
that has been analytically continued to negative $\omega$ via $\Gamma_\alpha(-\omega) = -\Gamma_\alpha(+\omega)$.
To obtain analytic results, we parametrize the latter by an ohmic form with a Lorentzian cutoff
\bea\label{EQ:gam_ohmic}
\Gamma_\alpha(\omega) = \Gamma_\alpha \frac{\omega}{\Omega}\frac{\delta_\alpha^2}{\omega^2+\delta_\alpha^2}\,,
\eea
where $\Gamma_\alpha$ encodes the system-reservoir coupling strength, $\delta_\alpha$ the cutoff width, 
and the energy scale $\Omega$ has just been introduced for dimensional convenience, such that $\Gamma_\alpha(\omega)$
has dimension of inverse time ($\hbar=1$ throughout).
We note that we expect the master equation description to hold when $\beta_\alpha \Gamma_\alpha \ll 1$.
This parametrization enables one to express the Lamb-shift in closed form as
\bea
\sigma_\alpha(\omega) &=& -\frac{\ii}{2} \frac{\delta_\alpha}{\omega} \Gamma_\alpha(\omega)
+ \frac{\ii}{2} \Gamma_\alpha(\omega) \cot\left(\frac{\beta_\alpha\delta_\alpha}{2}\right)\\
&&-\frac{\ii}{2\pi} \Gamma_\alpha(\omega) \left[\Psi\left(1+\frac{\beta_\alpha \delta_\alpha}{2\pi}\right)+\Psi\left(1-\frac{\beta_\alpha \delta_\alpha}{2\pi}\right)\right]\nn
&&+\frac{\ii}{2\pi} \Gamma_\alpha(\omega) \left[\Psi\left(1+\ii\frac{\beta_\alpha \omega}{2\pi}\right)+\Psi\left(1-\ii\frac{\beta_\alpha \omega}{2\pi}\right)\right]\,,\nonumber
\eea
where $\Psi(x)$ denotes the Polygamma (digamma) function~\cite{abramowitz1970}.
We note that $\sigma_\alpha(\omega)$ is purely imaginary.

The above quantum-optical master equation has many favorable properties:
First, due to its Lindblad form it preserves all density matrix properties, in particular positivity.
Second, we note that for the thermal reservoirs considered here the correlation functions satisfy 
Kubo-Martin-Schwinger (KMS) conditions~\cite{schaller2014}
\bea
\gamma_\alpha(-\omega) = e^{-\beta_\alpha \omega} \gamma_\alpha(+\omega)\,.
\eea
In last consequence, these relations can be used to show that for coupling to a 
single reservoir, a stationary state of the system is the thermalized one $\rho \propto e^{-\beta \HS}$ with
the inverse reservoir temperature $\beta$.
Third, it is also easy to see that when the spectrum of $\HS$ is non-degenerate, the populations (diagonals)
of the density matrix only couple to themselves
\bea
\dot{\rho}_{aa} = \sum_b \gamma_{ab,ab} \rho_{bb} - \left(\sum_b \gamma_{ba,ba}\right) \rho_{aa}
\eea
and do thus constitute a rate equation obeying local detailed balance properties.
When the spectrum of $\HS$ is partially degenerate, only coherences (off-diagonals) of the density matrix
corresponding to degenerate energies will couple to the remaining populations.
Depending on the application, the last properties may yield a tremendous reduction of the system dimensionality.


\subsection{Heat transport}

Consistently, we will consider off-diagonal matrix elements of the density matrix only when they
correspond to states that are energetically degenerate, i.e., $\rho_{ij}$ where $E_i=E_j$.
When there are no degeneracies in the spectrum of $\HS$, this implies that the master equation becomes a simple rate equation.
The rates $\gamma_{ia,jb}^\alpha$ from $\rho_{ab}$ to $\rho_{ij}$ will therefore only be non-vanishing when $E_a=E_b$ (since $E_i=E_j$ by construction), 
and can therefore be associated with the injection or extraction of energy $E_a-E_i$ from the system due to reservoir $\alpha$.
In the long-term limit, the system density matrix will assume a stationary value $\rho\to\bar\rho$, and the corresponding
energy current into the drain becomes
\bea\label{EQ:cur_eng}
I_E = \sum_{ab} \sum_i (E_a-E_i) \gamma_{ia,ib}^D \bar\rho_{ab}\,.
\eea
We note here that for a consistent thermodynamic description the energy current between system and reservoir should be
defined as above~\cite{wichterich2007a}.
The associated particle (matter) current is simply given by
\bea\label{EQ:cur_mat}
I_M = \sum_{ab} \sum_i \gamma_{ia,ib}^D \bar\rho_{ab}\,.
\eea
To determine the currents, one has to solve the master equation for the stationary state, 
i.e., in the energy eigenbasis one has to solve the large linear system
\bea
0 &=& -\ii (E_i-E_j) \bar\rho_{ij} -\ii \sum_a(\sigma_{ia}\bar\rho_{aj}-\sigma_{aj}\bar\rho_{ia})\nn
&&+\sum_{ab} \gamma_{ia,jb} \bar\rho_{ab} - \frac{1}{2} (\gamma_{ab,ai} \bar\rho_{bj}+\gamma_{aj,ab} \bar\rho_{ib})\,.
\eea
In contrast to our previous setup~\cite{vogl2012b} this has to be done numerically.
We note that the associated Liouvillians can become quite large, such that they can no longer
be stored in the computer memory as normal matrices.
Fortunately, the Liouvillians are also quite sparse, and by storing them in a sparse format
simple methods such as power iteration can be used to determine their stationary state.

\begin{figure}[ht]
\begin{center}
\includegraphics[width=0.48\textwidth,clip=true]{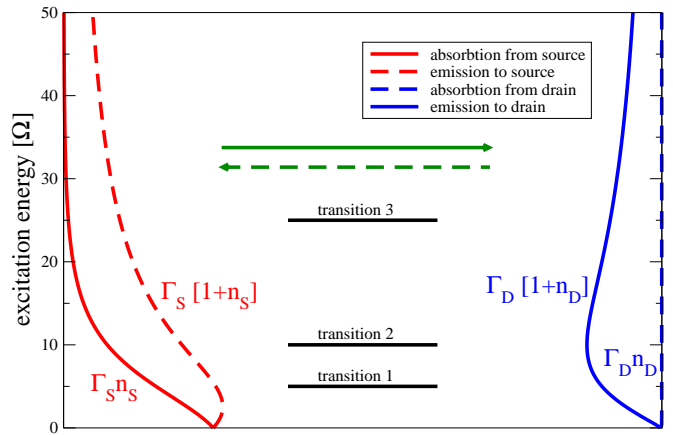}
\end{center}
\caption{\label{FIG:transport_spectroscopy}(Color Online)
Comparison of emission and absorption rates of source (left, red) and drain (right, blue) with
the transition frequencies of the system (center, black).
To support transport from source to drain, both the rates for absorption from the source (solid red) and emission into
the drain (solid blue) should be large.
For the shown large-bias configuration ($\beta_S \Omega = 0.1$, $\beta_D\Omega=10$) the current from source to drain (solid curves)
strongly exceeds the opposite current (dashed curves).
For the shown setup, the lowest excitation most strongly contributes to the heat current from source to drain.
Other parameters: $\Gamma_S=\Gamma_D$, $\delta_\alpha = 10 \Omega$.
}
\end{figure}

We note that although not directly evident from Eq.~(\ref{EQ:cur_eng}) and~(\ref{EQ:cur_mat}), the
net currents will vanish when e.g. the source is decoupled ($\Gamma_S(\omega) \to 0$), which
is enforced by the modified solution of the stationary state $\bar\rho_{ab}$.
The observation that the transition rate from state $a$ to state $i$ obeys 
$\gamma_{ia,ia}^\alpha\propto \Gamma_\alpha(E_a-E_i)[1+n_\alpha(E_a-E_i)]$ 
leads to a simple interpretation of the transition energies of the system, see Fig.~\ref{FIG:transport_spectroscopy}.
To enable for energetic transmissions through the system it is necessary that the rates at the corresponding transition frequency $\omega$
are finite both for absorption of energy from the source $\Gamma_S(\omega) n_S(\omega)$ and for emission to the drain 
$\Gamma_D(\omega)[1+n_D(\omega)]$. 
The situation in Fig.~\ref{FIG:transport_spectroscopy} is such that -- since the inverse transmission process has much smaller rates --
transport is in average directed from source to drain.
Furthermore, it is also visible that the third transition hardly contributes to transport.
The width of the transport window can be modified by the inverse temperatures $\beta_\alpha$ and
the widths of the spectral coupling densities $\delta_\alpha$.


\subsection{Spin chain diagonalization}

We will first discuss the analytically treatable homogeneous case ($g_i=g$ and $J^y_i=J^y$ and $J^z_i=J^z$)
with periodic boundary conditions.
Then, we discuss the numerically treatable non-homogeneous case of an open spin chain, 
where the homogeneous open spin chain appears as a special case.


\subsubsection{Diagonalization of the homogeneous periodic YZ model}

To diagonalize the YZ model in a transverse field
\bea\label{EQ:hamxyclosed}
\HS &=& - g \sum_{i=1}^N \sigma^x_i - J_y\sum_{i=1}^N \sigma^y_i \sigma^y_{i+1}
- J_z \sum_{i=1}^N \sigma^z_i \sigma^z_{i+1}\,,
\eea
we first for convenience introduce rescaled variables 
\bea
g = \Omega(1-s)\,,\quad
J_y = \Omega s \frac{1-\sigma}{2}\,,\quad
J_z = \Omega s \frac{1+\sigma}{2}\,,
\eea
where the parameter $\Omega$ denotes an energy scale of the system, to which all eigenvalues of
the system Hamiltonian are proportional.
The other parameters $s$ and $\sigma$ are dimensionless, and
we find the quantum phase transition from a paramagnetic phase to 
the ferromagnetic phases at $s_{\rm crit}=1/2$, the two ferromagnetic phases
are separated at $\sigma_{\rm crit}=0$ (anisotropy transition).

To avoid lengthy case distinctions we assume that the length $N$ of the chain is
an even number.
Inserting the Jordan-Wigner transform~\cite{jordan1928a} (see appendix \ref{APP:jordan_wigner}), the discrete Fourier
transform compatible with antiperiodic boundary conditions (see appendix \ref{APP:dft})
and the standard Bogoliubov transform (see appendix \ref{APP:bogoliubov_homogeneous})
one can now -- in the subspace of an even number of quasiparticles -- 
map the system Hamiltonian to non-interacting fermions~\cite{sachdev2000,dziarmaga2005a}
\bea\label{EQ:ham_isingdiag}
\HS = \sum_k \epsilon_k \left(\eta_k^\dagger \eta_k - \frac{1}{2}\right)\,,
\eea
where the 
energies read~\cite{bunder1999a}
\bea\label{Eeps}
\epsilon_k &=& 2 \Omega \Big[(1-s)^2+s^2 \left(\frac{1-\sigma}{2}\right)^2+s^2 \left(\frac{1+\sigma}{2}\right)^2\\
&&-2 s (1-s) \cos\left(\frac{2 \pi k}{N}\right)
+s^2\frac{1-\sigma^2}{2} \cos\left(\frac{4 \pi k}{N}\right)\Big]^{1/2}\,.\nonumber
\eea
Here, the quasimomentum $k$ may assume half-integer values 
\bea
k \in \left\{-\frac{N-1}{2}, -\frac{N-3}{2}, \ldots, +\frac{N-3}{2}, +\frac{N-1}{2}\right\}\,.
\eea
Closer inspection of the energies in Eq.~(\ref{Eeps}) yields that $\epsilon_k \ge 0$.

First, we note that the quasiparticle energies are symmetric $\epsilon_{+k} = \epsilon_{-k}$, which implies that some excited states
of the model are degenerate (e.g. states with in total two quasiparticles with quasimomenta $k_1=\pm 1/2$ and $k_2=\pm 3/2$).

Second, we also express the coupling operator in terms of the fermions
\bea\label{EQ:coupling_inhomog}
\sigma^x_n &=& \f{1} - \frac{2}{N} \sum_{kk'} \Big[\\
&&
+u_k^* u_{k'} \eta_{k}^\dagger \eta_{k'} e^{-\ii(k-k') \frac{2\pi n}{N}}
+v_k v_{k'}^* \eta_{k} \eta_{k'}^\dagger e^{+\ii(k-k') \frac{2\pi n}{N}}\nn
&&
+u_k^* v_{k'}^* \eta_k^\dagger \eta_{k'}^\dagger e^{-\ii(k+k') \frac{2\pi n}{N}}
+v_k u_{k'} \eta_k \eta_{k'} e^{+\ii(k+k') \frac{2\pi n}{N}}\Big]\nonumber
\eea
where it becomes obvious that these local couplings preserve the subspace of even quasiparticle numbers, 
as either only pairs of quasiparticles are created/annihilated or the total number of quasiparticles is
not changed at all.
Later-on, we will be particularly interested in the cases $n=N$ and $n=N/2$.
Here, the coefficients $u_k$ and $v_k$ are given by
\bea\label{EQ:ukvk}
u_k &\propto& \left[(1-s) - s \cos\left(\frac{2 \pi k}{N}\right) + \epsilon_k/(2\Omega)\right]\nn
v_k &\propto& s \sigma \sin\left(\frac{2\pi k}{N}\right)
\eea
with the normalization condition $\abs{u_k}^2+\abs{v_k}^2=1$.
Since $\epsilon_k\ge 0$ this implies that these coefficients can be chosen real.

Another particularly simple case arises for the coupling $J^x = \sum_i \sigma^x_i$, where we can collapse the summation
\bea\label{EQ:coupling_homog}
J^x &=& N \f{1} - 2 \sum_k \Big[\abs{u_k}^2 \eta_k^\dagger \eta_k + \abs{v_k}^2 \eta_k \eta_k^\dagger\nn
&&-u_{k}^* v_{k}^* \eta_{+k}^\dagger \eta_{-k}^\dagger - u_k v_{k} \eta_{-k} \eta_{+k}\Big]\,,
\eea
which formally coincides with our previous finding~\cite{vogl2012b} but now also includes the phase parameter $\sigma$ in the
coefficients $u_k$, $v_k$, and $\epsilon_k$.

We note here that it is straightforward to represent the system energy density $\expval{E}/N$ at finite temperature 
in the continuum limit $N\to\infty$.
Using that for a thermal state $\expval{E} = - \partial_\beta \ln Z_\beta$ with the partition function
$Z_\beta=\trace{e^{-\beta \HS}}$ with Hamiltonian~(\ref{EQ:ham_isingdiag}) one arrives at the expression
\bea
\frac{\expval{E}}{N} \to -\int\limits_0^{1/2} \epsilon(\kappa) \tanh\left[\frac{\beta\epsilon(\kappa)}{2}\right]d\kappa\,,
\eea
where $\epsilon(\kappa) = \epsilon_{(\kappa \cdot N)}$, compare Eq.~(\ref{Eeps}).
At zero temperature, this just becomes the ground state energy density which reflects the quantum criticality by a divergent
second derivative with respect to the control parameters $s$ and $\sigma$.
At finite temperature, critical dependence on these control parameters is no longer found.
Furthermore, we stress that the one-dimensional quantum Ising model has no thermal phase transition: 
The specific heat capacity (per spin, we use $k_{\rm B}=1$)
\bea
C = \frac{\partial}{\partial T} \left(\frac{\expval{E}}{N}\right) 
= \frac{1}{T^2} \int\limits_0^{1/2} \frac{\epsilon^2(\kappa)}{1+\cosh\left[\epsilon(\kappa)/T\right]} d\kappa
\eea
for example is an analytic function of temperature $T$.


\subsubsection{Diagonalization of the inhomogeneous open YZ model}

We consider the open YZ-model in a transverse field as a system
\bea\label{EQ:hamxy_open}
\HS &=& \sum_{i=1}^N g_i \sigma^x_i +\sum_{i=1}^{N-1} J^y_i \sigma^y_i \sigma^y_{i+1}
+ \sum_{i=1}^{N-1} J^z_i \sigma^z_i \sigma^z_{i+1}\,,
\eea
where we have considered open boundary conditions ($J^{y/z}_N=0$).

The Jordan-Wigner-transform (see appendix \ref{APP:jordan_wigner})
non-locally maps the Pauli spin matrices to fermionic annihilation and creation operators.
In particular, the resulting Hamiltonian is quadratic in the fermionic operators, which
means that it can be diagonalized with a general Bogoliubov transformation
\bea\label{EQ:bogoliubov_inhomog}
c_i = \sum_j \left(\alpha_{ij} \eta_j + \beta_{ij} \eta_j^\dagger\right)\,,
\eea
with new fermionic annihilation (creation) operators $\eta_j$ ($\eta_j^\dagger$) 
and complex-valued coefficients $\alpha_{ij}$ and $\beta_{ij}$.
Finding the optimal transformation can be mapped to the diagonalization of a $2N\times 2N$ matrix, 
see appendix~\ref{APP:bogoliubov_inhomog}.
The advantage of this procedure is that the complexity of obtaining the eigenvalues and eigenvectors is polynomial in the 
chain length $N$ and not exponential as a naive treatment would suggest.
Having solved the eigenvalue problem numerically (using e.g. a standard exact diagonalization routine~\cite{press1994}), the Hamiltonian 
can be represented in terms of the coefficients $\alpha_{ij}$ and $\beta_{ij}$
\bea\label{EQ:hamdisorderedchain}
H_S &=& \bar\epsilon \f{1} + \sum_j \epsilon_j \eta_j^\dagger \eta_j\,,\nn
\bar\epsilon &=& \sum_i g_i + \sum_{ij}\Big[-2 g_i \abs{\beta_{ij}}^2\nn
&&+ (J_i^y-J_i^z) (\alpha_{ij} \beta_{i+1,j} + {\rm h.c.})\nn
&&+ (J_i^y+J_i^z)(\beta_{ij} \beta_{i+1,j}^* + {\rm h.c.})\Big]\,,\nn
\epsilon_j &=& \sum_i \Big[
-2 g_i \abs{\alpha_{ij}}^2 + 2 g_i \abs{\beta_{ij}}^2\nn 
&&+ (J_i^y-J_i^z) (\alpha_{i+1,j} \beta_{ij} - \alpha_{ij} \beta_{i+1,j} + {\rm h.c.})\nn
&&+ (J_i^y+J_i^z) (\alpha_{ij} \alpha_{i+1,j}^* - \beta_{ij} \beta_{i+1,j}^* + {\rm h.c.})\Big]\,.
\eea
Obviously, both the vacuum energy $\bar\epsilon$ and the single-particle energies $\epsilon_j$ are real.
Furthermore, it is possible to choose the vacuum as the ground state and $\epsilon_j>0$.
Technically, the spectrum of a Hamiltonian with this form can easily be 
calculated in the Fock space representation:
Placing $n_i \in \{0,1\}$ quasiparticles in every mode $1\le i \le N$, the eigenvectors are
given by $\ket{\f{n}}=\ket{n_1, n_2, \ldots, n_N}$, and we have
$H \ket{\f{n}} = \left(\alpha + \sum_j \epsilon_j n_j\right) \ket{\f{n}}$.

A local coupling is in the fermionic quasiparticle basis represented as
\bea\label{EQ:coupling_chain}
\sigma^x_n &=& \f{1} - 2 \sum_{kj} \Big( \alpha_{nj}^* \alpha_{nk} \eta_j^\dagger \eta_k + \beta_{nj}^* \beta_{nk} \eta_j \eta_k^\dagger\nn
&&+ \alpha_{nj}^* \beta_{nk} \eta_j^\dagger \eta_k^\dagger + \beta_{nj}^* \alpha_{nk} \eta_j \eta_k\Big)\,,
\eea
and as before it is visible that only pairs of quasiparticles are created or annihilated.
For the open chain we will be naturally interested in the cases $n=1$ and $n=N$.


\subsection{Low-Temperature Limit}

Provided the ground state and the first excited state in the accessible Hilbert 
space are non-degenerate and sufficiently
far separated from the rest of the spectrum
we can at low temperatures $(E_i-E_1)\beta_\alpha \gg 1$
simplify the long-term dynamics of the resulting master equation to a $2\times 2$ rate 
equation containing only the ground and first excited state
\bea
{\cal L} &=& \sum_\alpha \Gamma_\alpha^{eg} M_\alpha^{eg}
\left(\begin{array}{cc}
-n_\alpha^{eg} & 1+n_\alpha^{eg}\\
n_\alpha^{eg} & -1-n_\alpha^{eg}
\end{array}\right)\,,
\eea
where $\Gamma_\alpha^{eg} = \Gamma_\alpha(E_1-E_0)$ is the bare emission/absorption rate
and $n_\alpha^{eg} = n_\alpha(E_1-E_0)$ the Bose distribution, both evaluated at the energy
gap between ground ($g$)and first excited ($e$) state.
The matrix element $M_\alpha^{eg} = \abs{\bra{0} A_\alpha \ket{1}}^2$ describes how efficient the
system coupling operators $A_{S/D}$ couple ground and first excited state.
The matter current then reduces to
\bea
I_M &=& \frac{\Gamma_S^{eg} \Gamma_D^{eg} M_S^{eg} M_D^{eg}}
{\Gamma_S^{eg} M_S^{eg}\left(1+2n_S^{eg}\right) + \Gamma_D^{eg}M_D^{eg}\left(1+2n_D^{eg}\right)}\times\nn
&&\times
 \left(n_S^{eg}-n_D^{eg}\right)\,,
\eea
whereas the energy current is tightly coupled to the matter current $I_E = (E_1-E_0) I_M$.
When we neglect the effect of asymmetric source-drain couplings $\Gamma_S^{eg} = \Gamma_D^{eg} = \Gamma$ and
in addition consistently expand the current for low-temperatures, it further reduces to
\bea\label{EQ:matelaverage}
I_M &\to& \Gamma \frac{M_S^{eg} M_D^{eg}}{M_S^{eg} + M_D^{eg}} \left(n_S^{eg}-n_D^{eg}\right)\nn
&\equiv& \Gamma M^{eg}\left(n_S^{eg}-n_D^{eg}\right)\,.
\eea
Thus, the scaling behaviour depends in the low-temperature limit strongly on the 
matrix elements $M_\alpha^{eg}$, and a first general idea about the dependence of the current
on system properties such as the chain length $N$ and disorder strength at low temperatures 
can be gained from analyzing $M^{eg}$.


\subsection{Consistency Checks}

Since the involved master equation are high-dimensional, it is essential to check
their consistency.
We have done this with a variety of tests (in case of numerical tests this is of course
limited by numerical accuracy):
First of all, positivity was preserved throughout due to the Lindblad type of the
used master equations.
Second, all currents vanished in equilibrium when $\beta_S=\beta_D=\beta$.
Third, in equilibrium we also found that the stationary density matrix was just given by
the thermal state in the accessible Hilbert space.
Fourth, for systems that were intrinsically symmetric between source and drain reservoirs 
(i.e., no disorder present) we found that the current changed sign when $\beta_S$ and $\beta_D$
were exchanged.
Fifth, for selected small-scale rate equation examples we verified the fluctuation theorem for 
heat exchange~\cite{saito2007b,harbola2007a}
\bea\label{EQ:fluctuation_theorem}
\lim_{t\to\infty} \frac{P_{+E}}{P_{-E}} = e^{(\beta_D-\beta_S) E}\,,
\eea
where $P_{+E}$ denotes the probability of an energy transfer $E$ from source to drain.
Technically, we verified this by testing the symmetry of the long-term cumulant-generating function 
\bea
{\cal C}(-\xi) = {\cal C}(+\xi + \ii (\beta_D-\beta_S))\,,
\eea
where $\xi$ denotes an energy counting field such that $P_E = \int e^{{\cal C}(\xi)} e^{-\ii E \xi} d\xi$.
For rate equations describing energy transport with two terminals and local detailed balance this is to be 
expected universally~\cite{andrieux2009a,campisi2011a}.
In particular, the exponent in the right-hand side of Eq.~(\ref{EQ:fluctuation_theorem}) will not depend on the 
microscopic properties of $\HS$.
Sixth, for the open spin chain we verified that the stationary current vanished when the ferromagnetic next-neighbor
interactions were turned off at any site of the chain -- effectively disconnecting source and drain reservoirs.
The same was trivially found true when one reservoir was disconnected.
Finally, we did of course verify that in the appropriate limits our master equation approached simplified analytic results.


\section{Perpendicular heat transport through the closed chain}\label{SEC:results1}

We consider the setup depicted in Fig.~\ref{FIG:coupling} a.) with
coupling operators $A_S=A_D=J^x$ and system Hamiltonian~(\ref{EQ:hamxyclosed}).
A closer look at the fermionic representation of the coupling operator in Eq.~(\ref{EQ:coupling_homog}) 
reveals that only the subspace of quasiparticle pairs with opposite quasimomenta will participate 
in the dynamics connected to the ground state.
This subspace -- of reduced dimension $2^{N/2}$ instead of $2^N$ is in general non-degenerate, 
and if degeneracies occur the corresponding states are not
connected by creation or annihilation of quasiparticle pairs.
Therefore, in this reduced subspace, a simple rate equation description applies.
In general, high-dimensional rate equations cannot be solved analytically.
This is particularly true for systems that do not have special structure such as e.g. tri-diagonal form.
Here however we have the particular case that source and drain coupling operators of the system are identical.
This leads to a special structure of the rate matrix
\bea
{\cal L} = \sum_\omega \sum_\alpha \Gamma_\alpha(\omega) {\cal L}_0^\omega 
+ \Gamma_\alpha(\omega) n_\alpha(\omega) {\cal L}_1^\omega\,,
\eea
where $\omega$ denotes the energy differences of the system, $\alpha\in\{S,D\}$ labels the reservoir with
bare emission/absorption rate $\Gamma_\alpha(\omega)$ and Bose distribution $n_\alpha(\omega)$, and
where ${\cal L}_{0/1}^\omega$ are superoperators that are independent of the reservoir.
Local detailed balance now implies that one has for all $\omega$ separately
\bea
\Gamma_\alpha(\omega) \left[{\cal L}_0^\omega + n_\alpha(\omega) {\cal L}_1^\omega\right] \bar\rho^\alpha = 0\,,
\eea
when $\bar{\rho}^\alpha$ is the thermal state, i.e., when
\bea
\frac{\bar{\rho}^\alpha_{n+1}}{\bar{\rho}^\alpha_n} = \frac{n_\alpha(E_{n+1}-E_n)}{1+n_\alpha(E_{n+1}-E_n)}\,.
\eea
The stationary state is thus determined by the function $n_\alpha(\omega)$, and by simply rewriting the 
equation for the total stationary state we obtain
\bea
0 &=& \sum_\omega \left[\Gamma_S(\omega)+\Gamma_D(\omega)\right]\times\nn
&&\times
\left[{\cal L}_0^\omega + \frac{\Gamma_S(\omega) n_S(\omega) + \Gamma_D(\omega) n_D(\omega)}{\Gamma_S(\omega)+\Gamma_D(\omega)} {\cal L}_1^\omega\right]\bar\rho\,,
\eea
such that the stationary state obeys
\bea\label{EQ:statstate}
\frac{\bar{\rho}_{n+1}}{\bar{\rho}_n} = \frac{\bar{n}(E_{n+1}-E_n)}{1+\bar{n}(E_{n+1}-E_n)}
\eea
with the average occupation
\bea
\bar{n}(\omega) = \frac{\Gamma_S(\omega) n_S(\omega) + \Gamma_D(\omega) n_D(\omega)}{\Gamma_S(\omega) + \Gamma_D(\omega)}\,.
\eea
We note that resulting stationary state~(\ref{EQ:statstate}) is in general -- 
unless the system has only a single transition frequency~\cite{vogl2011a} -- a non-thermal 
non-equilibrium steady state. This enables the calculation of
steady state expectation values such as the current, see below.


\subsection{Finite Thermal Bias}

The formal equivalence with our previous results~\cite{vogl2012b} allows us
to directly evaluate the stationary energy current 
\bea
I_E &=& \sum_{k>0} \frac{(2\epsilon_k) (4 u_k v_k)^2 \Gamma_S^k \Gamma_D^k (n_S^k - n_D^k)}
{\Gamma_S^k(1+2n_S^k)+\Gamma_D^k(1+2n_D^k)}\,,
\eea
where we have used the abbreviations $\Gamma_\alpha^k = \Gamma_\alpha(2\epsilon_k)$ and
$n_\alpha^k=n_\alpha(2\epsilon_k)$.
The expression for the matter current simply lacks the $(2\epsilon_k)$ factor.

Obviously, the current vanishes in equilibrium when $n_S(\omega) = n_D(\omega)$ as one would reasonably expect.
The current is suppressed when either source or drain are decoupled ($\Gamma_S^k=0$ or $\Gamma_D^k=0$, respectively).
When the coupling strength to source and drain is highly asymmetric, the smaller tunneling rate will bound the current 
(bottleneck).
Therefore, the maximum current can be expected when both reservoirs are about equally coupled but
have maximum difference in occupations $n_S^k \gg n_D^k$.

Most important however, we find that the current carries the quantum-critical signatures of the ground state somewhat
stronger than local observables such as magnetization or energy densities: Whereas the non-analyticities in the 
continuum limit only occur at strictly zero temperature, the current displays these even far from equilibrium, which we
discuss only for the infinite thermal bias limit below.


\subsection{Infinite thermal bias}

\begin{figure}[ht]
  \begin{overpic}[clip=true,width=\linewidth]{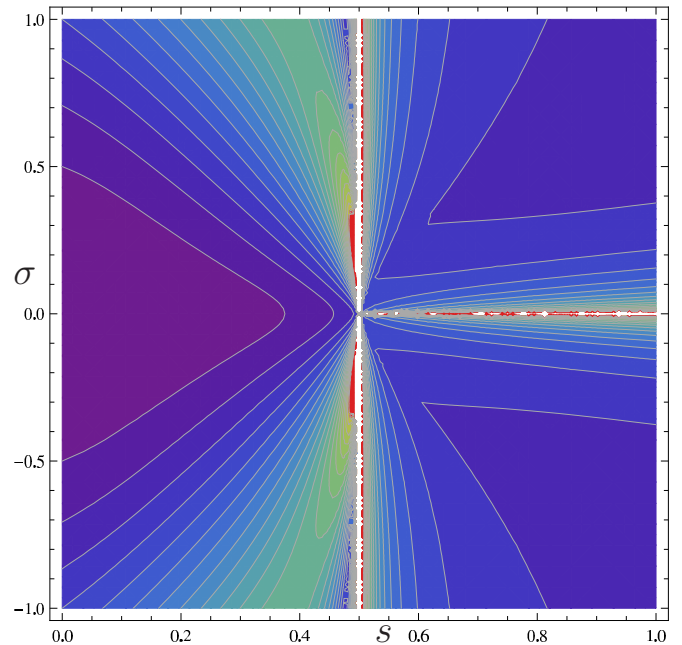}
    \put(0,55){\Large $\sigma$}
    \put(55,0){\Large $s$}
    \end{overpic}
\caption{\label{FIG:homog_xy}(Color Online)
Contour plot of the norm of the Hessian of the energy current (contours represent integer steps in units of $N \Omega \Gamma_D$, compare Eq.~(\ref{EQ:defhessian}) and Eq.~(\ref{EQ:ibcurrent_analytic})) 
in the limit of infinite thermal bias versus 
dimensionless phase parameters $s$ (horizontal axis) and $\sigma$ (vertical axis).
The critical lines (white) at $s=1/2$ and $\sigma=0$ are clearly visible as a divergent second derivative, 
and the current thus faithfully maps the phase diagram of the model.
The left sector describes the paramagnetic phase and the top (bottom) right sector the ferromagnetic phase in $z$ ($y$) direction.
}
\end{figure}

We will now focus on the infinite thermal bias limit ($n_S^k \to \infty$ and $n_D^k \to 0$) with flat
spectral coupling densities $\Gamma_\alpha^k \to \Gamma_\alpha$.
Provided that the source is not decoupled ($\Gamma_S>0$) we see that then the energy current is bounded by the rate to the drain only
\bea\label{EQ:ibcurrent_analytic}
I_E^\infty &=& \Gamma_D \sum_{k>0} (2 \epsilon_k) 8 u_k^2 v_k^2\\
&=& \sum_{k>0} \frac{8\Gamma_D \Omega s^2 \sigma^2 \sin^2\left(\frac{2\pi k}{N}\right)}
{\sqrt{s^2 \sigma^2 \sin^2\left(\frac{2\pi k}{N}\right) + \left(s+s\cos\left(\frac{2\pi k}{N}\right)-1\right)^2}}\nn
&\to& \frac{N \Gamma_D \Omega}{\pi} 4 s^2 \sigma^2\times\nn
&&\times
\int\limits_0^\pi 
\frac{\sin^2\left(\kappa\right)}
{\sqrt{s^2 \sigma^2 \sin^2\left(\kappa\right) + \left(s+s\cos\left(\kappa\right)-1\right)^2}} d\kappa\,,\nonumber
\eea
where in the last line we have assumed the continuum limit $N\to\infty$ to convert the summation into an integral.
The infinite bias limit leads to the fact that the source tunneling rate has no effect on the current: The
system essentially equilibrates with the hot source and is thus always maximally loaded, such that energy transfer
to the drain is only limited by $\Gamma_D$.
Obviously, the current vanishes at the lines defined by $s=0$ and $\sigma=0$: This is consistent with the fact that 
in these cases we have $\left[J^x,\HS\right]=0$, such that system and reservoirs cannot exchange energy.
Nevertheless, the second derivative of the infinite bias current faithfully detects the quantum-critical behaviour of the ground state 
also at the critical line $\sigma=0$.
To visualize the phase diagram of the model, we therefore plot the norm of the Hessian matrix
\bea\label{EQ:defhessian}
H \equiv \frac{1}{N \Omega \Gamma_D} \left(\begin{array}{cc}
\partial^2_s I_E^\infty & \partial_s \partial_\sigma I_E^\infty\\
\partial_\sigma \partial_s I_E^\infty & \partial^2_\sigma I_E^\infty
\end{array}\right)
\eea
as a function of the phase parameters $s$ and $\sigma$ in Fig.~\ref{FIG:homog_xy}.
It is visible that this norm diverges just at the critical lines of the model defined by $s_{\rm crit}=1/2$ and -- in the
ferromagnetic phase -- by $\sigma_{\rm crit}=0$.

For fixed $\sigma$ or $s$ a closed representation of the current in Eq.~(\ref{EQ:ibcurrent_analytic})
in terms of elliptic functions is possible (not shown).
Closer inspection of these analytic results reveals that near the critical lines the second derivatives of the 
current diverge logarithmically -- similar to the ground state energy density, e.g.
\bea
\left.\frac{\partial^2 I_E^\infty}{\partial s^2}\right|_{\sigma=1} &\propto& \ln\abs{s-\frac{1}{2}}\nn
\left.\frac{\partial^2 I_E^\infty}{\partial \sigma^2}\right|_{s=1} &\propto& \Theta(\sigma)\ln \sigma\,.
\eea

We note that if we had assumed an ohmic spectral coupling density $\Gamma_\alpha(\omega)$ as in
Eq.~(\ref{EQ:gam_ohmic}), the energy current would not display non-analytic behavior.
These would then however be persistent in the matter current (which we will analyze in the following models).
A super-ohmic choice would result in non-analytic behaviours of the energy current in higher derivatives.


\section{Longitudinal heat transport along the closed ring}\label{SEC:results2}

We consider the setup depicted in Fig.~\ref{FIG:coupling} b.) with 
system Hamiltonian~(\ref{EQ:hamxyclosed}) and coupling operators $A_S=\sigma^x_{N/2}$
and $A_D=\sigma^x_N$, compare Eq.~(\ref{EQ:coupling_inhomog}).
This situation is significantly more complicated than the previous case.
Nevertheless, it is not necessary to solve the full $2^N$-dimensional problem.
First, we note that with proper ground state initialization 
one can constrain the analysis to the subspace of an even number of quasiparticles, since 
the coupling operators~(\ref{EQ:coupling_inhomog}) can only create or annihilate pairs of quasiparticles.
This only reduces the dimension to $2^{N-1}$.
Unfortunately, even within this subspace many states are degenerate: Due to the symmetry
$\epsilon_{-k}=\epsilon_{+k}$ e.g. states with two quasiparticles with different absolute
quasimomenta are four-fold degenerate.
For a coupling to a single spin at site $n$, it is possible to show that there exists the
conserved quantity
\bea\label{EQ:consquant}
Q_n = \sum_q e^{+\ii 4\pi \frac{n q}{N}} \eta_{-q}^\dagger \eta_{+q}
\eea
obeying $\left[\sigma^x_n, Q_n\right]=0$ and $\left[\HS,Q_n\right]=0$, 
see Appendix~\ref{APP:consquant}.
Noting that $Q_{n+N/2} = - Q_n$ implies that the same quantity is conserved when we consider
transport through antipodal points, i.e., 
$\left[A_S, Q_N\right] = 0$ and $\left[A_D, Q_N\right]=0$.
The ground state (with no quasiparticles at all) has an associated eigenvalue $Q_0=0$, and
it therefore suffices to consider only those states that belong to the same subspace as the ground state.
Technically, this can be performed by an additional Bogoliubov transformation, see Appendix~\ref{APP:bogoliubov_add}.
Unfortunately, even in this subspace two-fold degeneracies remain, and the corresponding 
coherences have to be taken into account, which leads to an overall unfavorable scaling behaviour of 
the numerical effort with the ring length $N$, summarized in Table~\ref{TAB:scaling1}.
\begin{table}
\begin{tabular}{c|c|c|c}
length $N$ & dimension $d_H$ & dimension $d_L$ & entries $n_L$\\
\hline
4 & 6 & 8 & 56\\
6 & 20 & 32 & 416\\
8 & 70 & 142 & 2894\\
10 & 252 & 652 & 19192\\
12 & 924 & 3024 & 121488\\
14 & 3432 & 14016 & 738432\\
16 & 12870 & 64614 & 4331622\\
18 & 48620 & 295724 & 24629528\\
20 & 184756 & 1343056 & 136266416
\end{tabular}
\caption{\label{TAB:scaling1}
Scaling of relevant Hilbert space dimension $d_H$, Liouville dimension $d_L\times d_L$, and the number of non-vanishing
entries in the Liouvillian matrix $n_L$ with the system size $N$ for longitudinal 
heat transport through the closed Ising chain.
The sparsity of the Liouvillian can be used to calculate its stationary state.
}
\end{table}
Fortunately, the Liouvillian is sparse, as indicated by its number of non-vanishing entries $n_L$, such that
sparse matrix techniques have been used for storage and propagation. 
Furthermore, when one is interested only in e.g. the infinite thermal bias limit, the Liouvillian need not be explicitly
stored.
In this case, the dimension of the accessible Hilbert space determines the numerical effort:
Considering all possible transitions towards a basis state one can use that the coupling allows
to create or annihilate only two quasiparticles.
With a suitable numerical implementation, this corresponds to two bitflips, such that the 
numerical effort then only scales roughly as $N^2 \times d_H$.

\subsection{Finite temperature configuration}

\begin{figure}[ht]
\includegraphics[width=0.48\textwidth,clip=true]{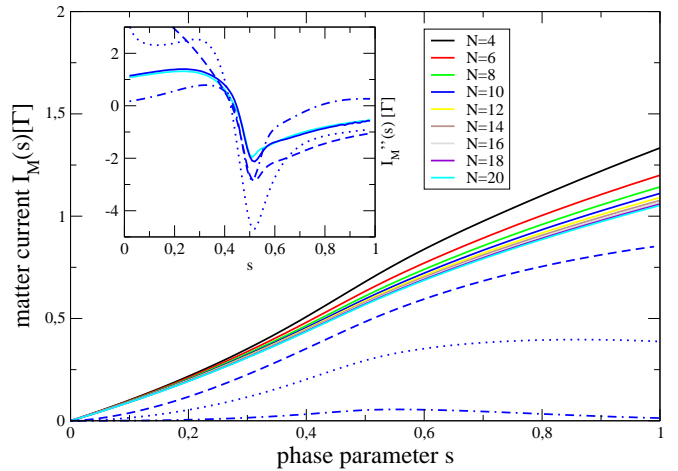}
\caption{(Color Online)\label{FIG:mattercurrent_ring}
Plot of the matter current through a ring for local couplings at $\sigma=1$ versus
phase parameter $s$.
At $s=1/2$ the ground state of the system Hamiltonian undergoes the finite-size analog of
a second-order quantum phase transition, which manifests itself as a dip in the second
derivative (inset).
In all curves, the temperature of the drain reservoir was negligible ($\gamma_D(\omega) \to \Gamma_D(\omega) \Theta(\omega)$).
Solid curves correspond to the infinite thermal bias configuration, the other curves (for $N=10$) denote high-bias (dashed, $\beta_S \Omega = 0.01$), 
intermediate bias with (dotted, $\beta_S \Omega = 0.1$), and low-bias (dash-dotted, $\beta_S \Omega = 1.0$) setups, respectively 
Other Parameters: $\Gamma_S=\Gamma_D=\Gamma$, $\delta_S=\delta_D=1000\Omega$, and $\beta_D\Omega \to \infty$.
}
\end{figure}

To set up the master equation correctly we introduced an ohmic spectral coupling density~(\ref{EQ:gam_ohmic})
and computed also the Lamb-shift terms explicitly.
We have not been able to generally solve the master equation analytically for its stationary state.
For an ohmic spectral coupling density, we did not find critical dependence of the energy current but for the matter current, 
which we evaluated numerically for finite problem sizes up to ring lengths of $N=20$ qubits.
In finite-sized realizations all divergences are rendered finite, instead of a 
true divergence one observes a dip, compare Fig.~\ref{FIG:mattercurrent_ring}.

\subsection{Low-temperature configuration}

For the closed spin chain with local couplings, the matrix elements of the coupling operators~(\ref{EQ:coupling_inhomog})
become independent of the coupling site in the low-temperature limit, such that the average
coupling in Eq.~(\ref{EQ:matelaverage}) becomes
\bea
M^{eg} = \abs{u_{1/2}}^2 \abs{v_{1/2}}^2 \frac{2}{N^2}\,,
\eea
with $u_{1/2}$ and $v_{1/2}$ obeying Eq.~(\ref{EQ:ukvk}).
Far from the critical point this scales inversely $I_M\propto 1/N^4$ with the chain length, 
and at the critical point one still observes a scaling as $I_M \propto 1/N^2$ (although here
the level separation condition cannot hold for large $N$, since ground state and first excited state
approach as $1/N$).
An inverse scaling with the chain length might be somewhat expected as the separation between
the reservoir increases with $N$. 
In contrast however, at infinite thermal bias we observe that the current becomes independent of the chain length.

\subsection{Infinite thermal bias}

In the limit of an infinite thermal bias ($n_D(\omega) \to \Theta(\omega)$ and $n_S(\omega) \to \infty$) one can however
achieve some simplifications.
First, we note that the Liouvillian of the source will dominate $\left|\left|L_S\right|\right| \gg \left|\left|L_D\right|\right|$, 
such that we can neglect the influence of the drain reservoir on the stationary state.
Using further that the source reservoir obeys local detailed balance relations we can conclude that the infinite bias stationary state 
is the high-temperature state $\rho_{ij} = \delta_{ij}/{\cal N}_{\rm red}$.
Here, ${\cal N}_{\rm red}$ is the number of states that participate in the dynamics, i.e., the total number of 
states with an even number of quasiparticles and with eigenvalue $\lambda_Q=0$ for the conserved quantity~(\ref{EQ:consquant}) 
with $n=N$.
Furthermore, the Bose distribution of the drain behaves for low temperatures such that
$\gamma_D(\omega) = \Gamma_D(\omega) \Theta(\omega)$.
For large widths $\delta_D\to\infty$, this implies that the matter current into the drain becomes
\bea\label{EQ:current_infbias}
I_M = \sum_{i a} (E_a-E_i) \Theta(E_a-E_i) \abs{\bra{a} A_D \ket{i}}^2 \frac{\Gamma_D}{\Omega {\cal N}_{\rm red}}\qquad\,.
\eea
Again we note that at infinite thermal bias the drain rate $\Gamma_D$ bounds the current.
The operator $A_D=\sigma^x_N$ only supports the creation or annihilation of two quasiparticles
(which enables one to efficiently evaluate the double summation in the above equation) and
thus the difference $E_a-E_i = \pm \epsilon_{k_1}\pm \epsilon_{k_2}$ just corresponds to two
energies of the quasiparticles.

In case of the Ising model in the ferromagnetic phase ($s=1$ and $\sigma=1$), all excitation energies become
similar ($\epsilon_k=2\Omega$), and creation of two quasiparticles is the only way for the system to gain energy.
Therefore, the expression for the current simplifies in this case to
\bea
\bar{I}_M &\to& 
\sum_{ia} \frac{4\Gamma_D}{{\cal N}_{\rm red}} \left(\frac{2}{N}\right)^2 \sum_{kk'} \abs{u_{k'}}^2 \abs{v_{k}}^2\times\nn
&&\times\left[\cos\left(\frac{\pi}{2}(k+k')\right)+\sin\left(\frac{\pi}{2}(k-k')\right)\right]^2\times\nn
&&\times\abs{\bra{a} \bar\eta_{k'}^\dagger \bar\eta_{k}^\dagger\ket{i}}^2\,,
\eea
where we have inserted the additional Bogoliubov transformation relating $\eta_k$ and $\bar\eta_k$, see
Appendix~\ref{APP:bogoliubov_add}.
Further using that $\sum_i \ket{i} \bra{i} = \f{1}$ we can simplify
\bea
\bar{I}_M
&\to& \frac{4 \Gamma_D 4}{{\cal N}_{\rm red} N^2} \sum_{k k'} \abs{u_{k'}}^2 \abs{v_{k}}^2\times\nn
&&\times\left[\cos\left(\frac{\pi}{2}(k+k')\right)+\sin\left(\frac{\pi}{2}(k-k')\right)\right]^2\times\nn
&&\times \sum_a \bra{a} \bar\eta_k^\dagger \bar\eta_k \bar\eta_{k'}^\dagger \bar\eta_{k'} \ket{a}\nn
&=& \frac{4 \Gamma_D}{N^2} \left[\sum_{k} \abs{v_{k}}^2 \left[\cos(\pi k/2) + \sin(\pi k/2)\right]^2\right]\times\nn
&&\times\left[\sum_{k'} \abs{u_{k'}}^2 \left[\cos(\pi k'/2) - \sin(\pi k'/2)\right]^2\right]\nn
&\stackrel{N\to\infty}{\to}& \Gamma_D\,.
\eea
In the last step, we have performed the continuum limit by replacing the summation over $k$ by
analytically solvable integrals.
Before, we have used that for very large $N$ one has
$\sum_a \bra{a} \bar\eta_k^\dagger \bar\eta_k \bar\eta_{k'}^\dagger \bar\eta_{k'}\ket{a} \approx {\cal N}_{\rm red}/4$, 
and we can see in Fig.~\ref{FIG:mattercurrent_ring} that the actual finite-size current is at $s=1$ only slightly above $\Gamma_D$.

Interestingly, we note that the current becomes for large chain lengths independent of $N$.
Naively, one might also at infinite bias have expected an inverse scaling of the current $I \propto N^{-\alpha}$ 
with the ring length, since for large chains the reservoirs are farther apart from each other.
Such an inverse scaling is also found in complementary works in the literature~\cite{benenti2009a,sun2010a,znidaric2011a,li2011a},
where local Lindblad operators were assumed (which is applicable in the strong-coupling limit~\cite{breuer2002} between system 
and reservoir).
However, the above considerations also show that although the matrix elements between two particular states become 
smaller for longer chains, the mere number of states participating in transport compensates this effect.
We expect that this behaviour is only found in the weak-coupling limit, where the pointer basis is given by the
energy eigenbasis.


\section{Heat transport along the open chain}\label{SEC:results3}

Here, we investigated the heat current through a homogeneous chain of type~(\ref{EQ:hamxy_open})
that is coupled at its ends to a source reservoir via $A_S = \sigma^x_1$ and a drain reservoir
via $A_D = \sigma^x_N$, compare Fig.~\ref{FIG:coupling} c.).
We exploit that also for an open spin chain or in presence of disorder it
is possible to diagonalize the system with only moderate efforts.
When the YZ model is opened ($J_N^y=J_N^z=0$, we observe that already this
removes degeneracies in the energy eigenbasis (these only remain in exceptional points), 
such that a rate equation description is in principle applicable.
We therefore calculated the energy eigenbasis diagonalizing a $2N\times 2N$ auxiliary
matrix as described in Appendix.~\ref{APP:bogoliubov_inhomog}.
Afterwards, we computed the current resulting from the rate equation.
The impact of disorder was investigated by distributing the ferromagnetic interaction
strengths in $y$-direction uniformly in the interval $J_i^y \in [\bar{J}^y-\Delta J^y/2, \bar{J}^y+\Delta J^y/2]$.
We calculated the current for 100  random instances of the spin chain and calculated afterwards mean and variance.
Throughout this section, the remaining parameters were chosen as 
$g_i = g = -(1-s)\Omega$ and $J_i^z = J^z = -s \Omega$.

A rate equation representation allows a much simpler numerical implementation of
the Liouvillian, since all entries are by construction real.
However, since we can only exploit that the number of quasiparticles is 
always even, the size storage requirements for longitudinal transport along the chain -- see Table~\ref{TAB:scaling2} -- 
\begin{table}
\begin{tabular}{c|c|c|c}
length $N$ & dimension $d_H$ & dimension $d_L$ & entries $n_L$\\
\hline
4 & 8 & 8 & 56\\
6 & 32 & 32 & 512\\
8 & 128 & 128 & 3712\\
10 & 512 & 512 & 23552\\
12 & 2048 & 2048 & 137216\\
14 & 8192 & 8192 & 753664\\
16 & 32768 & 32768 & 3964928\\
18 & 131072 & 131072 & 20185088\\
20 & 524288 & 524288 & 100139008
\end{tabular}
\caption{\label{TAB:scaling2}
Scaling of relevant Hilbert space dimension $d_H$, Liouville dimension $d_L\times d_L$, and the number of non-vanishing
entries in the Liouvillian matrix $n_L$ with the system size $N$ for longitudinal heat transport along the open chain.
}
\end{table}
are comparable to the case of longitudinal heat transport through the ring (Table~\ref{TAB:scaling1}).
Just as there, sparse matrix methods can be applied to find the stationary state of the Liouvillian.
In the limit of an infinite thermal bias however, it is not even necessary to store the full Liouvillian, 
and the required numerical effort approximately scales as $N^2 \times d_H$. 

\subsection{Infinite bias results}

\begin{figure}[ht]
\includegraphics[width=0.48\textwidth,clip=true]{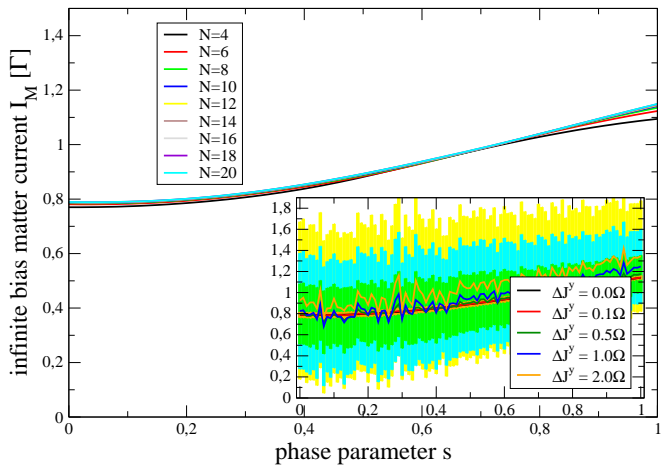}
\caption{\label{FIG:mattercurrent_chain}(Color Online)
Plot of the phonon current through the open spin chain with homogeneous parameters at infinite thermal bias.
Since $J^y=\Omega$ is finite throughout, the current does not vanish at $s=0$.
The inset demonstrates for $N=10$ that the effect of disorder 
(solid curves and lighter-shaded areas denote mean and standard deviation obtained from 100 random instances, respectively)
at infinite thermal bias is not drastic.
Other parameters: $\Gamma_S=\Gamma_D=\Gamma$, $J^x = -(1-s)\Omega$, $J^z = -s \Omega$, $J^y \in [\Omega-\Delta J^y/2, \Omega+\Delta J^y/2]$, 
$\delta_S=\Delta_D=1000\Omega$.
}
\end{figure}

Taking a closer look at the representation of the coupling operators in terms of the
fermionic quasiparticles~(\ref{EQ:coupling_chain}) we see that we can again constrain the dynamics
to the subspace of an even number of quasiparticles.
Using the local detailed balance properties of the rates we can conclude that the stationary
state in the infinite thermal bias regime will be the equipartitioned distribution of 
all states with an even number of quasiparticles, such that the infinite bias matter current
into the drain is formally identical to Eq.~(\ref{EQ:current_infbias}).
The result is depicted in Fig.~\ref{FIG:mattercurrent_chain}.
First, we observe that the infinite bias current is roughly independent of the chain length $N$:
Although the matrix element of the coupling operators between two energy eigenstates
decreases, the growing number of energy eigenstates compensates this effect.
In the inset we demonstrate the effect of increasing the disorder.
Somewhat surprisingly, it has little effect on the average current, only the width increases (shaded regions).

\subsection{Low-temperature limit}

\begin{figure}[ht]
\includegraphics[width=0.48\textwidth,clip=true]{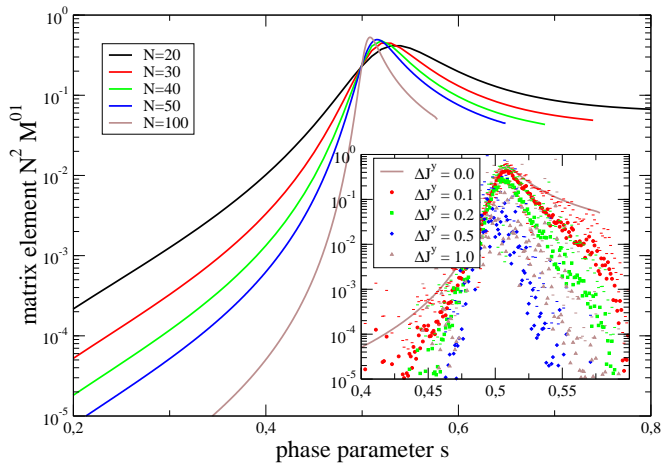}
\caption{\label{FIG:matelscaling}(Color Online)
Average coupling matrix element $M^{eg} = M_S^{eg} M_D^{eg}/(M_S^{eg}+M_D^{eg})$ constructed from Eq.~(\ref{EQ:matelseparate}) 
-- scaled by $N^2$ -- between ground state and first excited state (the energy and matter
current at low temperatures are essentially proportional to this coupling matrix element) versus phase parameter $s$.
The fact that all curves cross at the critical point is evidence that here the current scales as $\propto N^{-2}$ and decays
even faster further away from the critical point.
The inset shows the effect of disorder for $N=100$, where quite generally a suppression of the average current (symbols) is 
observed, especially far from the critical point.
Here, thin horizontal lines denote the standard deviation.
Other parameters: $J^x = -(1-s) \Omega$, $J^z=-s\Omega$, $J^y \in [-\Delta J^y/2, +\Delta J^y/2]$ (uniformly distributed).
}
\end{figure}

In the low-temperature limit and assuming level separation of the lowest two states, 
we investigated the impact of disorder on the current.
To do so, we chose the $J_i^y$ distributed uniformly in the interval $\bar{J}^y = [-\Delta J^y/2, +\Delta J^y/2]$, generated 
$100$ instances with $g_i = -(1-s)\Omega$ and $J_i^z = J^z = -s \Omega$, and calculated the
current for each instance in dependence on $s$.
For the open spin chain we obtain -- adopting the convention that 
$\epsilon_a < \epsilon_b < \epsilon_{j \neq \{a,b\}}$ are the smallest two
single-particle quasienergies in Eq.~(\ref{EQ:hamdisorderedchain}) -- the expression
\bea\label{EQ:matelseparate}
M_S^{eg} &=& 4 \left(\abs{\alpha_{1a}}^2 \abs{\beta_{1b}}^2 
+ \abs{\alpha_{1b}}^2 \abs{\beta_{1a}}^2\right)\,,\nn
M_D^{eg} &=& 4 \left(\abs{\alpha_{Na}}^2 \abs{\beta_{Nb}}^2 
+ \abs{\alpha_{Nb}}^2 \abs{\beta_{Na}}^2\right)\,,
\eea
where $\alpha_{ij}$ and $\beta_{ij}$ are determined by diagonalizing the matrix
Eq.~(\ref{EQ:diagmatrix}).
Afterwards, we calculated average and variance of the current.
As expected, the mean current scales inversely with the system size $N$, see Fig.~\ref{FIG:matelscaling}.
At the critical point, numerical data suggest a scaling $\propto N^{-2}$ as with the closed chain (compare the crossing point of 
all curves).
Furthermore, the inset shows that the presence of disorder further reduces
the current, which we interpret as due to the onset of localization of ground state
and first excited state, similar to Anderson localization~\cite{anderson1961a}.


\section{Scaling versus thermal bias}\label{SEC:fourier}

\begin{figure}[ht]
\includegraphics[width=0.48\textwidth,clip=true]{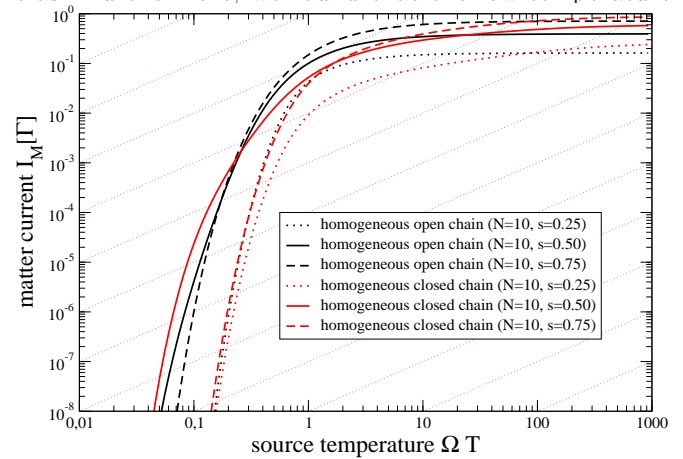}
\caption{\label{FIG:curversbias}(Color Online)
Plot of the phonon matter current versus temperature difference for the closed chain (black curves) and the open 
chain (red curves) for different regions in the phase space and assuming a vanishing drain temperature.
We note that at low source temperature the current scales inversely with the chain length $N$ -- with the exponent
depending on the phase.
At high temperature differences the current reaches a maximum that is independent of $N$.
Consistent with our previous results, at low bias the current through critical systems 
is strongly enhanced in comparison to non-critical system.
In contrast, at high bias this is no longer pronounced.
Thin straight lines in the background indicate the slope that would be observed if Fourier's law was obeyed.
Other parameters: $\Gamma_S=\Gamma_D=\Gamma$, $\delta_S=\delta_D=1000\Omega$, $J^x = -(1-s)\Omega$, $J^y=0$, $J^z=-s\Omega$, and $\beta_D\Omega\to\infty$.
}
\end{figure}

It is of course possible to combine the different scaling behaviours observed for longitudinal transport 
at small (inverse with $N$) and large (independent of $N$)  thermal biases, see Fig.~\ref{FIG:curversbias}.
It is visible that at low bias, the current strongly decays, and the conductance is surprisingly small.
In contrast, at infinite thermal bias the maximum conductance becomes independent on the chain length.

The varying slope of the curves and the fact that the current becomes independent
of the temperature difference at large thermal bias
indicates that Fourier's law $I \propto \Delta T$ is not satisfied by the 
systems considered.
We attribute this to the integrability of the models considered~\cite{saito1996a,saito2000a,mejia_monasterio2005a,manzano2012a,asadian2013a,li2011b},
which is also not broken if disorder is constrained to the discussed type.
Fourier's law is expected to hold for non-integrable models~\cite{saito2003a}.


\section{Conclusions}

We have calculated the heat transport through integrable spin models connected to two
bosonic reservoirs kept at different temperatures.

For the closed homogeneous YZ model with homogeneous coupling to the reservoirs we
were able to derive an analytic solution for the current.
We found that the quantum-critical behaviour of the ground state was faithfully mapped 
by the current even at extreme bias configurations far from equilibrium.
Reasonably, we found for the perpendicular heat transport a linear scaling of the current with
the spin chain length.

Complexity was strongly increased when we considered more realistic local couplings to the heat baths.
We could exploit the existence of conserved quantities depending on the coupling site to
reduce the dimension.
In the infinite bias regime we found that the current saturated independent of the chain length, 
whereas at small temperatures an inverse scaling could be demonstrated.
Nevertheless, also in this case our numerical results indicated that the current shows signatures
of the critical behaviour.

In the case of an open chain we could -- with a simple rate equation approach -- 
reproduce the saturation of the current for large chain lengths and infinite thermal bias.
Furthermore, we found that the low-temperature current also shows signatures of the critical point, which
remains true when weak disorder is present.
Remarkably, we found that disorder did not significantly reduce the average infinite bias current.

None of the systems obeyed Fourier's law of heat conduction, which we attribute to their integrability.
We would also like to emphasize that for a consistent thermodynamic description of transport one should
microscopically derive the master equation, since local terms in the interaction Hamiltonian will in 
general not map to local jump terms in the master equation.


\section{Acknowledgements}

Financial support by the DFG (SCHA 1646/2-1) is gratefully acknowledged.


\bibliographystyle{unsrt}
\bibliography{/home/schaller/literatur/postdoc/postdoc}

\begin{thebibliography}{10}

\bibitem{sachdev2000}
S.~Sachdev.
\newblock {\em Quantum Phase Transitions}.
\newblock Cambridge University Press, 2000.

\bibitem{cejnar2006a}
Pavel Cejnar, Michal Macek, Stefan Heinze, Jan Jolie, and Jan Dobeš.
\newblock Monodromy and excited-state quantum phase transitions in integrable
  systems: collective vibrations of nuclei.
\newblock {\em Journal of Physics A: Mathematical and General}, 39(31):L515,
  2006.

\bibitem{caprio2008a}
M.A. Caprio, P.~Cejnar, and F.~Iachello.
\newblock Excited state quantum phase transitions in many-body systems.
\newblock {\em Annals of Physics}, 323(5):1106 -- 1135, 2008.

\bibitem{perez_fernandez2011a}
P.~P\'erez-Fern\'andez, A.~Rela\~no, J.~M. Arias, P.~Cejnar, J.~Dukelsky, and
  J.~E. Garc\'ia-Ramos.
\newblock Excited-state phase transition and onset of chaos in quantum optical
  models.
\newblock {\em Phys. Rev. E}, 83:046208, Apr 2011.

\bibitem{dietz2013a}
B.~Dietz, F.~Iachello, M.~Miski-Oglu, N.~Pietralla, A.~Richter, L.~von Smekal,
  and J.~Wambach.
\newblock Lifshitz and excited-state quantum phase transitions in microwave
  dirac billiards.
\newblock {\em Phys. Rev. B}, 88:104101, Sep 2013.

\bibitem{brandes2013a}
Tobias Brandes.
\newblock Excited-state quantum phase transitions in dicke superradiance
  models.
\newblock {\em Phys. Rev. E}, 88:032133, 2013.

\bibitem{lignier2007a}
H.~Lignier, C.~Sias, D.~Ciampini, Y.~Singh, A.~Zenesini, O.~Morsch, and
  E.~Arimondo.
\newblock Dynamical control of matter-wave tunneling in periodic potentials.
\newblock {\em Phys. Rev. Lett.}, 99:220403, 2007.

\bibitem{baumann2010a}
K.~Baumann, C.~Guerlin, F.~Brennecke, and T.~Esslinger.
\newblock Dicke quantum phase transition with a superfluid gas in an optical
  cavity.
\newblock {\em Nature}, 464:1301, 2010.

\bibitem{baumann2011a}
K.~Baumann, R.~Mottl, F.~Brennecke, and T.~Esslinger.
\newblock Exploring symmetry breaking at the dicke quantum phase transition.
\newblock {\em Phys. Rev. Lett.}, 107:140402, Sep 2011.

\bibitem{inoue2010a}
Jun-ichi Inoue and Akihiro Tanaka.
\newblock Photoinduced transition between conventional and topological
  insulators in two-dimensional electronic systems.
\newblock {\em Phys. Rev. Lett.}, 105:017401, Jun 2010.

\bibitem{lindner2011a}
Netanel~H. Lindner, Gil Refael, and Victor Galitski.
\newblock Floquet topological insulator in semiconductor quantum wells.
\newblock {\em Nature Physics}, 7:490–495, 2011.

\bibitem{jiang2011a}
Liang Jiang, Takuya Kitagawa, Jason Alicea, A.~R. Akhmerov, David Pekker, Gil
  Refael, J.~Ignacio Cirac, Eugene Demler, Mikhail~D. Lukin, and Peter Zoller.
\newblock Majorana fermions in equilibrium and in driven cold-atom quantum
  wires.
\newblock {\em Phys. Rev. Lett.}, 106:220402, Jun 2011.

\bibitem{bastidas2012a}
V.~M. Bastidas, C.~Emary, B.~Regler, and T.~Brandes.
\newblock Nonequilibrium quantum phase transitions in the dicke model.
\newblock {\em Physical Review Letters}, 108:043003, 2012.

\bibitem{bastidas2012b}
V.~M. Bastidas, C.~Emary, G.~Schaller, and T.~Brandes.
\newblock Nonequilibrium quantum phase transitions in the ising model.
\newblock {\em Physical Review A}, 86:063627, 2012.

\bibitem{morrison2008a}
S.~Morrison and A.~S. Parkins.
\newblock Dynamical quantum phase transitions in the dissipative
  lipkin-meshkov-glick model with proposed realization in optical cavity qed.
\newblock {\em Phys. Rev. Lett.}, 100:040403, Jan 2008.

\bibitem{dalla_torre2010a}
Emanuele~G. {Dalla Torre}, Eugene Demler, Thierry Giamarchi, and Ehud Altman.
\newblock Quantum critical states and phase transitions in the presence of
  non-equilibrium noise.
\newblock {\em Nature Physics}, 6:806–810, 2010.

\bibitem{bhaseen2012a}
M.~J. Bhaseen, J.~Mayoh, B.~D. Simons, and J.~Keeling.
\newblock Dynamics of nonequilibrium dicke models.
\newblock {\em Phys. Rev. A}, 85:013817, Jan 2012.

\bibitem{kessler2012a}
E.~M. Kessler, G.~Giedke, A.~Imamoglu, S.~F. Yelin, M.~D. Lukin, and J.~I.
  Cirac.
\newblock Dissipative phase transition in a central spin system.
\newblock {\em Physical Review A}, 86:012116, Jul 2012.

\bibitem{hoening2012a}
M.~H\"oning, M.~Moos, and M.~Fleischhauer.
\newblock Critical exponents of steady-state phase transitions in fermionic
  lattice models.
\newblock {\em Phys. Rev. A}, 86:013606, Jul 2012.

\bibitem{lambert2009a}
Neill Lambert, Yueh-nan Chen, Robert Johansson, and Franco Nori.
\newblock Quantum chaos and critical behavior on a chip.
\newblock {\em Physical Review B}, 80:165308, Oct 2009.

\bibitem{vogl2012b}
Malte Vogl, Gernot Schaller, and Tobias Brandes.
\newblock Criticality in transport through the quantum ising chain.
\newblock {\em Physical Review Letters}, 109:240402, 2012.

\bibitem{esposito2010a}
Massimiliano Esposito and Christian Van~den Broeck.
\newblock Three faces of the second law. i. master equation formulation.
\newblock {\em Physical Review E}, 82(1):011143, 2010.

\bibitem{campisi2011a}
Michele Campisi, Peter H\"anggi, and Peter Talkner.
\newblock Colloquium: Quantum fluctuation relations: Foundations and
  applications.
\newblock {\em Rev. Mod. Phys.}, 83(3):771--791, 2011.

\bibitem{lindblad1976a}
G.~Lindblad.
\newblock On the generators of quantum dynamical semigroups.
\newblock {\em Communications in Mathematical Physics}, 48:119--130, 1976.

\bibitem{friedenauer2008a}
A.~Friedenauer, H.~Schmitz, J.~T. Glueckert, D.~Porras, and T.~Schaetz.
\newblock Simulating a quantum magnet with trapped ions.
\newblock {\em Nature Physics}, 4:757, 2008.

\bibitem{mostame2008a}
Sarah Mostame and Ralf Sch\"utzhold.
\newblock Quantum simulator for the ising model with electrons floating on a
  helium film.
\newblock {\em Physical Review Letters}, 101:220501, 2008.

\bibitem{zhang2009a}
Jingfu Zhang, Fernando~M. Cucchietti, C.~M. Chandrashekar, Martin Laforest,
  Colm~A. Ryan, Michael Ditty, Adam Hubbard, John~K. Gamble, and Raymond
  Laflamme.
\newblock Direct observation of quantum criticality in ising spin chains.
\newblock {\em Physical Review A}, 79:012305, Jan 2009.

\bibitem{coldea2010a}
R.~Coldea, D.~A. Tennant, E.~M. Wheeler, E.~Wawrzynska, D.~Prabhakaran,
  M.~Telling, K.~Habicht, P.~Smeibidl, and K.~Kiefer.
\newblock Quantum criticality in an ising chain: Experimental evidence for
  emergent $e_8$ symmetry.
\newblock {\em Science}, 327:177, 2010.

\bibitem{edwards2010a}
E.~E. Edwards, S.~Korenblit, K.~Kim, R.~Islam, M.-S. Chang, J.~K. Freericks,
  G.-D. Lin, L.-M. Duan, and C.~Monroe.
\newblock Quantum simulation and phase diagram of the transverse-field ising
  model with three atomic spins.
\newblock {\em Physical Review B}, 82:060412, Aug 2010.

\bibitem{kim2011b}
K~Kim, S~Korenblit, R~Islam, E~E Edwards, M-S Chang, C~Noh, H~Carmichael, G-D
  Lin, L-M Duan, C~C~Joseph Wang, J~K Freericks, and C~Monroe.
\newblock Quantum simulation of the transverse ising model with trapped ions.
\newblock {\em New Journal of Physics}, 13(10):105003, 2011.

\bibitem{bunder1999a}
J.~E. Bunder and R.~H. McKenzie.
\newblock Effect of disorder on quantum phase transitions in anisotropic xy
  spin chains in a transverse field.
\newblock {\em Physical Review B}, 60:344 -- 358, 1999.

\bibitem{pfeuty1970a}
P.~Pfeuty.
\newblock The one-dimensional ising model with a transverse field.
\newblock {\em Annals of Physics}, 57:79--90, 1970.

\bibitem{arrachea2009a}
Liliana Arrachea, Gustavo~S. Lozano, and A.~A. Aligia.
\newblock Thermal transport in one-dimensional spin heterostructures.
\newblock {\em Phys. Rev. B}, 80:014425, 2009.

\bibitem{saito2003a}
K.~Saito.
\newblock Strong evidence of normal heat conduction in a one-dimensional
  quantum system.
\newblock {\em Europhysics Letters}, 61:34–40, 2003.

\bibitem{prosen2010a}
Tomaž Prosen and Bojan Zunkovič.
\newblock Exact solution of markovian master equations for quadratic fermi
  systems: thermal baths, open xy spin chains and non-equilibrium phase
  transition.
\newblock {\em New Journal of Physics}, 12(2):025016, 2010.

\bibitem{sun2010a}
Ke-Wei Sun, Chen Wang, and Qing-Hu Chen.
\newblock Heat transport in an open transverse-field ising chain.
\newblock {\em EPL (Europhysics Letters)}, 92(2):24002, 2010.

\bibitem{li2011b}
Wenjuan Li and Peiqing Tong.
\newblock Heat conduction in one-dimensional aperiodic quantum ising chains.
\newblock {\em Phys. Rev. E}, 83:031128, Mar 2011.

\bibitem{breuer2002}
H.-P. Breuer and F.~Petruccione.
\newblock {\em The Theory of Open Quantum Systems}.
\newblock Oxford University Press, Oxford, 2002.

\bibitem{schaller2008a}
G.~Schaller and T.~Brandes.
\newblock Preservation of positivity by dynamical coarse-graining.
\newblock {\em Physical Review A}, 78:022106, 2008.

\bibitem{schaller2014}
Gernot Schaller.
\newblock {\em Open Quantum Systems Far from Equilibrium}.
\newblock Springer, 2014.

\bibitem{abramowitz1970}
Milton Abramowitz and Irene~A. Stegun, editors.
\newblock {\em Handbook of Mathematical Functions}.
\newblock National Bureau of Standards, 1970.

\bibitem{wichterich2007a}
Hannu Wichterich, Markus~J. Henrich, Heinz-Peter Breuer, Jochen Gemmer, and
  Mathias Michel.
\newblock Modeling heat transport through completely positive maps.
\newblock {\em Physical Review E}, 76(3):031115, 2007.

\bibitem{jordan1928a}
P.~Jordan and E.~Wigner.
\newblock \"uber das paulische \"aquivalenzverbot.
\newblock {\em Zeitschrift f\"ur Physik}, 47:631--651, 1928.

\bibitem{dziarmaga2005a}
Jacek Dziarmaga.
\newblock Dynamics of a quantum phase transition: Exact solution of the quantum
  ising model.
\newblock {\em Physical Review Letters}, 95:245701, 2005.

\bibitem{press1994}
William~H. Press, Saul~A. Teukolsky, William~T. Vetterling, and Brian~P.
  Flannery.
\newblock {\em Numerical Recipes in C}.
\newblock Cambridge University Press, $2^{\rm nd}$ edition, 1994.

\bibitem{saito2007b}
Keiji Saito and Abhishek Dhar.
\newblock Fluctuation theorem in quantum heat conduction.
\newblock {\em Physical Review Letters}, 99(18):180601, 2007.

\bibitem{harbola2007a}
Upendra Harbola, Massimiliano Esposito, and Shaul Mukamel.
\newblock Statistics and fluctuation theorem for boson and fermion transport
  through mesoscopic junctions.
\newblock {\em Physical Review B}, 76(8):085408, Aug 2007.

\bibitem{andrieux2009a}
D.~Andrieux, P.~Gaspard, T.~Monnai, and S.~Tasaki.
\newblock The fluctuation theorem for currents in open quantum systems.
\newblock {\em New Journal of Physics}, 11:043014, 2009.

\bibitem{vogl2011a}
M.~Vogl, G.~Schaller, and T.~Brandes.
\newblock Counting statistics of collective photon transmissions.
\newblock {\em Annals of Physics}, 326:2827, 2011.

\bibitem{benenti2009a}
G.~Benenti, G.~Casati, T.~Prosen, D.~Rossini, and M.~\v{Z}nidari\v{c}.
\newblock Charge and spin transport in strongly correlated one-dimensional
  quantum systems driven far from equilibrium.
\newblock {\em Physical Review B}, 80(3):035110, 2009.

\bibitem{znidaric2011a}
M.~Znidaric.
\newblock Transport in a one-dimensional isotropic heisenberg model at high
  temperature.
\newblock {\em Journal of Statistical Mechanics: Theory and Experiment},
  2011(12):P12008, 2011.

\bibitem{li2011a}
Jun Li, Yu~Liu, Jing Ping, Shu-Shen Li, Xin-Qi Li, and YiJing Yan.
\newblock Large-deviation analysis for counting statistics in mesoscopic
  transport.
\newblock {\em Physical Review B}, 84:115319, 2011.

\bibitem{anderson1961a}
P.~W. Anderson.
\newblock Localized magnetic states in metals.
\newblock {\em Physical Review}, 124:41 -- 53, 1961.

\bibitem{saito1996a}
K.~Saito, S.~Takesue, and S.~Miyashita.
\newblock Thermal conduction in a quantum system.
\newblock {\em Phys. Rev. E}, 54:2404--2408, 1996.

\bibitem{saito2000a}
K.~Saito, S.~Takesue, and S.~Miyashita.
\newblock Energy transport in the integrable system in contact with various
  types of phonon reservoirs.
\newblock {\em Phys. Rev. E}, 61:2397--2409, Mar 2000.

\bibitem{mejia_monasterio2005a}
T.~Prosen C.~Mejia-Monasterio and G.~Casati.
\newblock Fourier's law in a quantum spin chain and the onset of quantum chaos.
\newblock {\em Europhysics Letters}, 72:520–526, 2005.

\bibitem{manzano2012a}
Daniel Manzano, Markus Tiersch, Ali Asadian, and Hans~J. Briegel.
\newblock Quantum transport efficiency and fourier's law.
\newblock {\em Phys. Rev. E}, 86:061118, Dec 2012.

\bibitem{asadian2013a}
A.~Asadian, D.~Manzano, M.~Tiersch, and H.~J. Briegel.
\newblock Heat transport through lattices of quantum harmonic oscillators in
  arbitrary dimensions.
\newblock {\em Phys. Rev. E}, 87:012109, Jan 2013.

\end{thebibliography}


\newpage

\appendix


\section{Jordan-Wigner transform}\label{APP:jordan_wigner}

The Jordan-Wigner-transform is given by
\bea
\sigma^x_n &=& \f{1} - 2 c_n^\dagger c_n\,,\nn
\sigma^y_n &=& -\ii \left(c_n - c_n^\dagger\right) \prod_{m=1}^{n-1} \left(\f{1} - 2 c_m^\dagger c_m\right)\,,\nn
\sigma^z_n &=& - \left(c_n + c_n^\dagger\right) \prod_{m=1}^{n-1} \left(\f{1} - 2 c_m^\dagger c_m\right)\,.
\eea
It non-locally maps the Pauli spin matrices to fermionic annihilation and creation operators, where the 
anti-commutation relations can be easily checked.
The inverse of the Jordan-Wigner transform is therefore given by
\bea
c_n &=& - \frac{1}{2} \left(\sigma^z_n - \ii \sigma^y_n\right) \prod_{m=1}^{n-1} \sigma^x_m\,,\nn
c_n^\dagger &=& - \frac{1}{2} \left(\sigma^z_n + \ii \sigma^y_n\right) \prod_{m=1}^{n-1} \sigma^x_m\,.
\eea


\section{Discrete Fourier Transform}\label{APP:dft}

The discrete Fourier transform
\bea
c_n &=& \frac{e^{-\ii \pi/4}}{\sqrt{N}} \sum_k \tilde{c}_k e^{+\ii k n \frac{2\pi}{N}}\,,\nn
c_n^\dagger &=& \frac{e^{+\ii \pi/4}}{\sqrt{N}} \sum_k \tilde{c}_k^\dagger e^{-\ii k n \frac{2\pi}{N}}
\eea
maps fermionic operators to new fermionic operators.
Its inverse is readily given by
\bea
\tilde{c}_k &=& \frac{e^{+\ii \pi/4}}{\sqrt{N}} \sum_n c_n e^{-\ii k n \frac{2\pi}{N}}\,,\nn
\tilde{c}_k^\dagger &=& \frac{e^{-\ii \pi/4}}{\sqrt{N}} \sum_n c_n^\dagger e^{+\ii k n \frac{2\pi}{N}}\,.
\eea
In these formulas, the quasimomentum takes the values $k\in\{\pm\frac{1}{2},\pm \frac{3}{2}, \ldots, \pm \frac{N-1}{2}\}$,
whereas the sites are labeled as $n\in\{1,2,\ldots,N\}$.


\section{Homogeneous Bogoliubov Transform}\label{APP:bogoliubov_homogeneous}

The Bogoliubov transform
\bea
\tilde{c}_k &=& u_{+k} \eta_{+k} + v_{-k}^* \eta_{-k}^\dagger\,,\nn
\tilde{c}_k^\dagger &=& u_{+k}^* \eta_{+k}^\dagger + v_{-k} \eta_{-k}
\eea
maps fermions to fermions and preserves the anti-commutation relations when 
$u_{+k} v_{+k}^* + u_{-k} v_{-k}^* = 0$ and $\abs{u_{+k}}^2 + \abs{v_{-k}}^2 = 1$.
The specific choice of the coefficients
\bea
u_k &\propto& \left[(1-s) - s \cos\left(\frac{2 \pi k}{N}\right) + \epsilon_k/(2\Omega)\right]\nn
v_k &\propto& s \sigma \sin\left(\frac{2\pi k}{N}\right)
\eea
with $\epsilon_k$ from Eq.~(\ref{Eeps}) shows that to
diagonalize the Hamiltonian, the coefficients can be chosen real.
In particular, we have the relations $u_{-k} = + u_{+k}$ and $v_{-k} = - v_{+k}$.
Computing the inverse transformation therefore yields
\bea
\eta_{+k} &=& u_{-k}^* \tilde{c}_{+k} - v_{-k}^* \tilde{c}_{-k}^\dagger\,,\nn
\eta_{+k}^\dagger &=& u_{-k} \tilde{c}_{+k}^\dagger - v_{-k} \tilde{c}_{-k}\,.
\eea


\section{Inhomogeneous Bogoliubov Transform}\label{APP:bogoliubov_inhomog}

After the Jordan-Wigner transformation, the Hamiltonian~(\ref{EQ:hamxy_open}) reads
\bea
\HS &=& \sum_{i=1}^N \left[g_i \f{1} - 2 g_i c_i^\dagger c_i \right]\nn
&&+ \sum_{i=1}^{N} \Big[
+ (J_i^y-J_i^z) \left(c_i c_{i+1} + c_{i+1}^\dagger c_i^\dagger\right)\nn
&&+ (J_i^y+J_i^z) \left(c_i^\dagger c_{i+1} + c_{i+1}^\dagger c_i\right)\Big]\,,
\eea
i.e., it is a quadratic fermionic Hamiltonian that can be diagonalized with standard methods.
To obtain a diagonal (single-particle) representation, where the hopping terms like $c_i c_{i+1}^\dagger$ and 
also particle-nonconserving terms like $c_i c_{i+1}$ are absent, we introduce the most general Bogoliubov transformation
(\ref{EQ:bogoliubov_inhomog}).
The new operators also obey fermionic anti-commutation relations
$\left\{\eta_i, \eta_j^\dagger\right\} = \delta_{ij}$ and $\left\{\eta_i, \eta_j\right\} = 0$.
Preservation of the anti-commutation relations requires for the coefficients
\bea\label{Econd1}
0 &=& \sum_j \left(\alpha_{ij} \beta_{i'j}+\alpha_{i'j} \beta_{ij}\right)\,,\nn
\delta_{ii'} &=& \sum_j \left(\alpha_{ij} \alpha_{i'j}^* + \beta_{ij} \beta_{i'j}^*\right)\,,
\eea
where $^*$ denotes the complex conjugate.
This does not yet fix the transformation. 
In addition, we demand that the Hamiltonian becomes diagonal in quasiparticle representation
\bea
H_S = \sum_j \left[\alpha_j \eta_j^\dagger \eta_j + \beta_j \eta_j \eta_j^\dagger + \gamma_j\f{1}\right]\,.
\eea
Inserting the Bogoliubov transformation and setting -- after appropriate symmetry transformations -- the coefficients of undesired terms to zero 
yields further constraints, which read explicitly
\bea\label{Econd2}
\delta_{jk} &\propto& 
\sum_i \Big[-2 g_i \alpha_{ij}^* \alpha_{ik} + 2 g_i \beta_{ij} \beta_{ik}^*\\
&&+(J_i^y-J_i^z)\times\nn
&&\times (\beta_{ij} \alpha_{i+1,k} + \alpha_{i+1,j}^* \beta_{ik}^* - \beta_{i+1,j} \alpha_{ik} - \alpha_{ij}^* \beta_{i+1,k}^*)\nn
&&+(J_i^y+J_i^z)\times\nn
&&\times (\alpha_{ij}^* \alpha_{i+1,k} + \alpha_{i+1,j}^* \alpha_{ik} - \beta_{i+1,j} \beta_{ik}^* - \beta_{ij} \beta_{i+1,k}^*)\Big]\nn
0 &=& -2 g_i \beta_{ij}^* \alpha_{ik} + 2 g_i \alpha_{ij} \beta_{ik}^*\nn
&&+(J_i^y-J_i^z)\times\nn
&&\times (\alpha_{ij} \alpha_{i+1,k} + \beta_{i+1,j}^* \beta_{ik}^* - \alpha_{i+1,j} \alpha_{ik} - \beta_{ij}^* \beta_{i+1,k}^*)\nn
&&+(J_i^y+J_i^z)\times\nn
&&\times (\beta_{ij}^* \alpha_{i+1,k} + \beta_{i+1,j}^* \alpha_{ik} - \alpha_{i+1,j} \beta_{ik}^* - \alpha_{ij} \beta_{i+1,k}^*)\Big]\nonumber
\eea
Solving Eqns.~(\ref{Econd1}) and~(\ref{Econd2}) simultaneously can be mapped to solving the eigenvalue 
problem of a hermitian matrix, such that a solution always exists.

First, by defining the $2N$ vectors
\bea
\ket{V_k} = \left(\begin{array}{c}
\left(\begin{array}{c}
\alpha_{k1}\\
\beta_{k1}
\end{array}\right)\\
\vdots\\
\left(\begin{array}{c}
\alpha_{kN}\\
\beta_{kN}
\end{array}\right)
\end{array}\right)\,,\qquad
\ket{\bar{V}_k} = \left(\begin{array}{c}
\left(\begin{array}{c}
\beta_{k1}^*\\
\alpha_{k1}^*
\end{array}\right)\\
\vdots\\
\left(\begin{array}{c}
\beta_{kN}^*\\
\alpha_{kN}^*
\end{array}\right)
\end{array}\right)\,,\qquad
\eea
we see that Eqns.(\ref{Econd1}) can be expressed as the orthonormality conditions
$\braket{\bar{V}_{k'}}{\bar{V}_k}=\delta_{kk'}$, 
$\braket{V_{k'}}{V_k}=\delta_{kk'}$, and
$\braket{\bar{V}_{k'}}{V_k}=0$.
Hence, arranging these vectors in a unitary matrix
\bea
{\cal U} = 
\left(\begin{array}{ccc}
\left(\begin{array}{cc}
\alpha_{11} & \beta_{11}^*\\
\beta_{11} & \alpha_{11}^*
\end{array}\right)
&
\hdots
&
\left(\begin{array}{cc}
\alpha_{N1} & \beta_{N1}^*\\
\beta_{N1} & \alpha_{N1}^*
\end{array}\right)\\
\vdots & & \vdots\\
\left(\begin{array}{cc}
\alpha_{1N} & \beta_{1N}^*\\
\beta_{1N} & \alpha_{1N}^*
\end{array}\right)
&
\hdots
&
\left(\begin{array}{cc}
\alpha_{NN} & \beta_{NN}^*\\
\beta_{NN} & \alpha_{NN}^*
\end{array}\right)
\end{array}\right)
\eea
we use that the rows are also by construction orthonormal to define the vectors
\bea
\ket{W_k} = \left(\begin{array}{c}
\left(\begin{array}{c}
\alpha_{1k}\\
\beta_{1k}^*
\end{array}\right)\\
\vdots\\
\left(\begin{array}{c}
\alpha_{Nk}\\
\beta_{Nk}^*
\end{array}\right)
\end{array}\right)\,,\qquad
\ket{\bar{W}_k} = \left(\begin{array}{c}
\left(\begin{array}{c}
\beta_{1k}\\
\alpha_{1k}^*
\end{array}\right)\\
\vdots\\
\left(\begin{array}{c}
\beta_{Nk}\\
\alpha_{Nk}^*
\end{array}\right)
\end{array}\right)\,.\qquad
\eea
Choosing these orthonormal $\braket{W_k}{W_{k'}} = \delta_{kk'}$, $\braket{\bar{W}_k}{\bar{W}_{k'}} = \delta_{kk'}$, and
$\braket{\bar{W}_k}{W_{k'}} = 0$ automatically satisfies Eqns.~(\ref{Econd1}).

One can show that Eqns.~(\ref{Econd2}) be written in terms of the vectors $\ket{W_k}$ and $\ket{\bar{W}_k}$ as
$\bra{W_j} M \ket{W_k} \propto \delta_{jk}$ or $-\bra{{\bar W}_j} M \ket{{\bar W}_k}^* \propto \delta_{jk}$, respectively, and
$\bra{\bar{W}_j} M \ket{W_k} = 0$.
The matrix $M$ is the 
$2N \times 2N$-dimensional hermitian and band diagonal matrix
\bea\label{EQ:diagmatrix}
M = \left(\begin{array}{ccccc}
\f{D}_1 & \f{J}_1 & \f{0} & \hdots &\f{0}\\
\f{J}_1^\dagger & \f{D}_2 & \f{J}_2 & \ddots &\vdots\\
\f{0} & \f{J}_2^\dagger & \f{D}_3 & \f{\ddots} &\f{0}\\
\vdots & \ddots & \ddots & \ddots &\f{J}_{N-1}\\
\f{0} & \hdots & \f{0} & \f{J}_{N-1}^\dagger & \f{D}_N
\end{array}\right)
\eea
with the $2\times 2$ matrices
\bea
\f{D}_i &=& \left(\begin{array}{cc}
-2 g_i & 0\\
0 & +2 g_i
\end{array}\right)\,,\nn
\f{J}_i &=& \left(\begin{array}{cc}
+(J_i^y+J_i^z) & -(J_i^y-J_i^z)\\
+(J_i^y-J_i^z) & -(J_i^y+J_i^z)
\end{array}\right)\,.
\eea
Clearly, Eqns.~(\ref{Econd2}) (and of course orthonormality~(\ref{Econd1})) are automatically fulfilled when one chooses 
the $\ket{W}$ and $\ket{\bar{W}}$ vectors as eigenvectors of the matrix $M$.
The matrix $M$ has a mirror symmetry in its eigenvalues, for each positive eigenvalue $M \ket{W_k} = +\epsilon_k \ket{W_k}$ we have
also the negative eigenvalue $M \ket{\bar{W}_k} = - \epsilon_k \ket{\bar{W}_k}$.


\section{Conserved Quantities for local couplings}\label{APP:consquant}

The system Hamiltonian has many degenerate levels (e.g. those where the total quasimomentum does not vanish).
This is due to the fact that only the absolute value of the quasimomentum $k$ enters the energy.
One can easily show that an operator of the form
$Q = \sum_{q} f_q \eta_{-q}^\dagger \eta_{+q}$
where hermiticity requires $f_{-q}^* = f_{+q}$, 
will commute with the system Hamiltonian.
Therefore, we use this as an ansatz to find a conserved quantity under local couplings.
To reduce the computational effort we first rewrite the coupling operator (\ref{EQ:coupling_inhomog}) as
\bea
\sigma^x_n &=& \left[1-\frac{2}{N} \sum_k \abs{v_k}^2\right]\f{1}\nn
&&-\frac{2}{N} \sum_{kk'} \left( u_k^* u_{k'} - v_k^* v_{k'}\right) e^{-\ii(k-k') \frac{2\pi n}{N}} \eta_k^\dagger \eta_{k'}\nn
&&-\frac{2}{N} \sum_{kk'} \left( v_k u_{k'} e^{+\ii(k+k') \frac{2\pi n}{N}} \eta_k \eta_{k'} + {\rm h.c.}\right)\,,
\eea
where the terms in the second row conserve the total quasiparticle number and the terms in the last row do not.
To obtain a conserved quantity in presence of the reservoir, the commutator of both terms $[Q,\sigma^x_n] = C_1+C_2$ should vanish.
The commutator arising from the particle-number conserving terms becomes
\bea
C_1 &=& \sum_{kk'q} f_q (u_{k}^* u_{k'}-v_{k}^* v_{k'}) e^{-\ii(k-k') \frac{2\pi n}{N}}\left[\eta_{-q}^\dagger \eta_{+q}, \eta_{+k}^\dagger \eta_{+k'}\right]\nn
&=& \sum_q \left(f_{-q} e^{+\ii \frac{4\pi n q}{N}} - f_{+q} e^{-\ii \frac{4\pi n q}{N}}\right)\eta_{+q}^\dagger\eta_{+q}\nn
&& +\sum_q \sum_{k\neq\pm q}\left(f_{-q} e^{+\ii(q+k) \frac{2\pi n}{N}} -f_{+k} e^{-\ii(q+k) \frac{2\pi n}{N}}\right)\times\nn
&&\times (u_q^* u_k + v_q^* v_k) \eta_q^\dagger \eta_k\nn
&=& \sum_{qk} \left(f_{-q} e^{+\ii(q+k) \frac{2\pi n}{N}}-f_{+k} e^{-\ii(q+k) \frac{2\pi n}{N}}\right)\times\nn
&&\times (u_q^* u_k + v_q^* v_k) \eta_q^\dagger \eta_k\,.
\eea
The second commutator becomes
\bea
C_2 &=& \sum_{kk'q} f_q v_k u_{k'} e^{+\ii(k+k')\frac{2\pi n}{N}} \left[\eta_{-q}^\dagger \eta_{+q}, \eta_{+k} \eta_{+k'}\right]\nn
&=& \sum_q \sum_{k\neq\pm q} \left(-f_q e^{+\ii(k-q)\frac{2\pi n}{N}} + f_k e^{-\ii(k-q) \frac{2\pi n}{N}}\right)\times\nn
&&\times u_k v_q \eta_k \eta_q\nn
&=& \sum_q \sum_{k\neq\pm q} \left(f_{-q} e^{+\ii(k+q)\frac{2\pi n}{N}}-f_{+k}e^{-\ii(k+q)\frac{2\pi n}{N}}\right)\times\nn
&&\times u_k v_q \eta_k \eta_{-q}
\eea
To obtain a conserved quantity, we have to demand that both $C_1$ and $C_2$ (the hermitian conjugate term does
not yield additional constraints) vanish, which yields a condition on $f_q$
\bea
f_{-q} e^{+\ii(q+k) \frac{2\pi n}{N}}=f_{+k} e^{-\ii(q+k) \frac{2\pi n}{N}}\,.
\eea
This is fulfilled by choosing
\bea
f_q = e^{\ii \frac{4\pi q n}{N}}\,,
\eea
such that the operator in Eq.~(\ref{EQ:consquant}) is a constant of motion
even in presence of a reservoir that couples to a site $n$ on the chain.


\section{Additional Bogoliubov transformation}\label{APP:bogoliubov_add}
The observation that there exists a conserved quantity for transport through antipodal points suggests to diagonalize the
Hamiltonian and that conserved quantity simultaneously.
We therefore use the non-standard Bogoliubov transformation
\bea
\eta_q = a_q \bar\eta_q + b_q \bar\eta_{-q}\,,\qquad
\eta_q^\dagger = a_q^* \bar\eta_q^\dagger + b_q^* \bar\eta_{-q}^\dagger
\eea
with coefficients satisfying $\abs{a_q}^2 + \abs{b_q}^2 =1$ and $a_q b_{-q}^* + b_q a_{-q}^* =0$ to preserve
the fermionic anti-commutation relations.
The Hamiltonian is left invariant when $\abs{a_q}^2+\abs{b_{-q}}^2=1$ and $a_q^* b_q + b_{-q}^* a_{-q}=0$.
Demanding that the conserved quantity is diagonal yields the equation
\bea
e^{+4\pi\ii \frac{n q}{N}} a_{-q}^* a_{+q} + e^{-4\pi\ii \frac{n q}{N}} b_{+q}^* b_{-q} = 0\,,
\eea
which implies that the sought-after transformation will depend on the site of the coupling $n$.
We however are seeking for a transformation that is identical for two coupling operators $n=N/2$ and $n=N$.
In this case, it is easy to show that all equations can be simultaneously fulfilled with
\bea
a_q = \frac{1}{\sqrt{2}} e^{\ii\left(\frac{\pi}{2} q + \frac{\pi}{4}\right)}\,,\qquad
b_q = \frac{1}{\sqrt{2}} e^{\ii\left(\frac{\pi}{2} q - \frac{\pi}{4}\right)}\,.
\eea
In the new representation, system Hamiltonian and the conserved quantity become
\bea
\HS &=& \sum_q \epsilon_q \left(\bar\eta_q^\dagger \bar\eta_q - \frac{1}{2}\right)\,,\nn
Q_{N/2}=-Q_N &=& \sum_q \sin(\pi q) \bar\eta_q^\dagger \bar\eta_q\,.
\eea
The coupling operators for coupling to site $n=N/2$ and $n=N$ become
\bea
A_{N/2} &=& \left(1-\frac{2}{N} \sum_k \abs{v_{+k}}^2\right) \f{1}\nn
&&-\frac{2}{N} \sum_{kk'} \Big[\left(u_{k}^* u_{k'} - v_{k}^* v_{k'}\right) \cos\left(\frac{3\pi}{2}(k-k')\right)\nn
&&- \left(u_{k}^* u_{k'} + v_{k}^* v_{k'}\right) \sin\left(\frac{3\pi}{2}(k+k')\right)\Big] \bar\eta_k^\dagger \bar\eta_{k'}\nn
&&-\frac{2}{N} \sum_{kk'} \Big\{\ii v_k u_{k'} \bar\eta_k \bar\eta_{k'}\times\nn
&&\times \left[\cos\left(\frac{3\pi}{2}(k+k')\right) + \sin\left(\frac{3\pi}{2}(k-k')\right)\right] + {\rm h.c.}\Big\}\,,\nn
A_N &=& \left(1-\frac{2}{N} \sum_k \abs{v_{+k}}^2\right) \f{1}\nn
&&-\frac{2}{N} \sum_{kk'} \Big[\left(u_{k}^* u_{k'} - v_{k}^* v_{k'}\right) \cos\left(\frac{\pi}{2}(k-k')\right)\nn
&&- \left(u_{k}^* u_{k'} + v_{k}^* v_{k'}\right) \sin\left(\frac{\pi}{2}(k+k')\right)\Big] \bar\eta_k^\dagger \bar\eta_{k'}\nn
&&-\frac{2}{N} \sum_{kk'} \Big\{\ii v_k u_{k'} \bar\eta_k \bar\eta_{k'}\times\nn
&&\times \left[\cos\left(\frac{\pi}{2}(k+k')\right) + \sin\left(\frac{\pi}{2}(k-k')\right)\right] + {\rm h.c.}\Big\}\,,
\eea
and closer inspection of the coefficients yields that $A_{N/2}$ and $A_N$ may partially trigger different transitions within the system.

\end{document}